\documentclass[aip,jcp,amsmath,amssymb,reprint,oneside]{revtex4-1}
\usepackage{graphicx}
\usepackage{dcolumn}
\usepackage{bm}
\usepackage{mathrsfs}
\usepackage{amsmath, amssymb, amsthm, amsfonts}
\usepackage[dvipsnames]{xcolor}
\usepackage{ulem}
\usepackage{caption}
\usepackage{hyperref}
\usepackage[textfont=rm]{subcaption}
\usepackage[textfont=rm]{subcaption}
\usepackage{soul}
\usepackage[percent]{overpic}
\usepackage{float}

\begin{document}

\title{Growth of Structural Lengthscale in Kob Andersen Binary Mixtures: Role of medium range order}

\author{Sanket Kumawat }
\affiliation{Polymer Science and Engineering Division, CSIR-National Chemical Laboratory, Pune-411008,
India}
\affiliation{Academy of Scientific and Innovative Research (AcSIR), Ghaziabad 201002, India}

\author{Mohit Sharma }
\affiliation{Polymer Science and Engineering Division, CSIR-National Chemical Laboratory, Pune-411008,
India}
\affiliation{Academy of Scientific and Innovative Research (AcSIR), Ghaziabad 201002, India}
\author{Ujjwal Kumar Nandi }
\affiliation{Polymer Science and Engineering Division, CSIR-National Chemical Laboratory, Pune-411008, India}
\affiliation{\textit{Department of Physics, Kyoto University, Kyoto 606-8502, Japan}}
\author{Indrajit Tah}
\affiliation{CSIR-Central Glass  Ceramic Research Institute, India}
\affiliation{Academy of Scientific and Innovative Research (AcSIR), Ghaziabad 201002, India}

\author{Sarika Maitra Bhattacharyya}
\email{mb.sarika.ncl@csir.res.in}

\affiliation{Polymer Science and Engineering Division, CSIR-National Chemical Laboratory, Pune-411008,
India}
\affiliation{Academy of Scientific and Innovative Research (AcSIR), Ghaziabad 201002, India}

\date{\today}

\begin{abstract}
A central and extensively debated question in glass physics concerns whether a single, growing lengthscale fundamentally controls glassy dynamics, particularly in systems lacking obvious structural motifs like the Kob Andersen binary Lennard Jones (KALJ) model.
In this work, we investigate structural and dynamical lengthscales in supercooled liquids using KALJ model in two compositions: 80:20 and 60:40. We compute the dynamical lengthscale from displacement–displacement correlation functions and observe a consistent growth as temperature decreases. To explore the static counterpart, we use a structural order parameter (SOP) based on the mean field caging potential. While this SOP is known to predict short time dynamics effectively, its bare correlation function reveals minimal spatial growth. Motivated by recent findings that long time dynamics reflect collective rearrangements, we perform spatial coarse-graining of the SOP and identify an optimal lengthscale $L_{max}$ that maximises structure–dynamics correlation. We show that the structural correlation length derived from SOP coarse-grained over $L_{max}$ exhibits clear growth with cooling and closely tracks the dynamical lengthscale, especially for A particles in the 80:20 mixture and for both A and B particles in the 60:40 system. Our results reconcile the previously observed absence of static length growth in the KALJ model by highlighting the necessity of intermediate range structural descriptors. Furthermore, we find that the particles with larger structural length growth also correspond to species with latent crystallisation tendencies, suggesting a possible link between structural order, dynamics, and incipient crystallisation. 

\end{abstract}

\maketitle
\section{Introduction}
\label{sec:introduction}

Supercooled liquids exhibit a dramatic slowdown in dynamics upon cooling and ultimately form an amorphous solid, or glass~\cite{Debenedetti2001,Cavagna2009,Berthier2011,Ediger2000}. 
Despite sustained progress, the microscopic mechanisms driving this glass transition remain debated. A central challenge is to rationalise the rapid increase in the dynamics, spanning many orders of magnitude, with a corresponding static structural signature \cite{tanakamrco_kawasaki_prl_2007, Coslovich2007,royall2008,tanakamrco_shintani_natphys_2008, Karmakar2009, tanakamrco_critical_natmat_2010, keys2011, Coslovich2011, Sastry2014, bapst2020, Manoj_prl_2021, Sharma2022}. 
This issue has motivated extensive efforts to identify growing structural and dynamical correlation lengths, a key theme in modern glass physics~\cite{shintani2006, tanakamrco_kawasaki_prl_2007, tanakamrco_critical_natmat_2010, Dunleavy2012,Tarjus_simpleglass_2012, MalinsLifetimes2013,russo_tanaka,Tah_PRL_2018}.

Dynamical lengthscales are relatively well established, often quantified through multi-point correlation functions such as the four-point dynamic structure factor \cite{Karmakar2010, Berthier2011, Tah_dis_dis, Paula2023} or the displacement–displacement correlation function ~\cite{Donati1999PRL,Tah_dis_dis,Paula2023}. Studies of dynamical lengthscale have also been extended in the deep supercooled region below the mode coupling theory (MCT) transition temperature\cite{Flenner2013,Hallett2018,berthier_PRX_2022}. In some of these studies, it was shown that the growth of the dynamical lengthscale slows down in this deep supercooled region\cite{Flenner2013,Hallett2018}.
 
Over the past two decades, considerable effort has also been devoted to identifying static lengthscales that grow alongside the well established dynamical correlation length in glass forming systems. 
Tanaka and coworkers have extensively studied the growth of static lengthscale in systems exhibiting MRCO, such as those with Steinhardt bond orientational order parameter\cite{Steinhardt1983} in 3D and hexatic order in 2D and found that these systems show a stronger coupling between static and dynamic lengthscales~\cite{tanakamrco_shintani_natphys_2008, tanakamrco_kawasaki_prl_2007, tanakamrco_kawasaki_JPCM_2010, tanakamrco_critical_natmat_2010, tanakamrco_leocmach_natcomm_2012, tanakamrco_boo_jncs_2012, tanakamrco_hu_pree_2016}.
 
Note that in 2D systems, the presence of MRCO, which is primarily a hexatic order, is quite common and easy to detect \cite{tanakamrco_critical_natmat_2010,russo_tanaka,tanakamrco_kawasaki_prl_2007,TongTanaka2018,TarjusPRL2010}. However, in 3D systems, the MRCO is not so common and is obtained mostly in deep supercooling. In marginally polydisperse systems, these structures are face centered cubic (fcc) \cite{tanakamrco_leocmach_natcomm_2012} or hexagonal close packing \cite{tanakamrco_critical_natmat_2010}  and in systems with higher polydispersity they are icosahedral or defective icosahedral structures \cite{Hallett2018}. Advanced techniques for measuring static lengthscales, such as the point-to-set (PTS) method, were designed to be order-agnostic \cite{biroli_thermodynamic_2008,Hocky2012,Biroli2013,Berthier2016}. However, Russo and Tanaka \cite{russo_tanaka} showed that for a two dimensional polydisperse system exhibiting strong hexatic order, the PTS lengthscale fails to capture the underlying hexatic order that actually drives the slowing down of the dynamics. They suggested that the PTS length primarily reflects two-body correlations encoded in the radial distribution function.
Tah \textit{et al.} \cite{Tah_PRL_2018} investigated the PTS length, distinguishing between those with and without MRCO. Their work for 2D systems revealed a sublinear scaling of the static PTS length with the dynamical correlation length for systems lacking MRCO, while for MRCO systems, these two lengths were found to grow commensurately if the cavity was chosen in a way that the inside liquid mimics the thermodynamic properties of the bulk liquid. Royall and co-workers have shown that for polydisperse systems in deep supercooling, the static lengthscale connected to the locally favoured structure, which is a defective icosahedra order, grows almost in a similar fashion as the dynamical lengthscale \cite{Hallett2018}. 

However, for MRCO-free systems like the 3D Kob Andersen binary Lennard Jones mixture, this correspondence often breaks down. Early studies by Marinari and Pitard \cite {Marinari2005} and La Nave \textit{et al.} \cite{LaNave2006} linked cooperative relaxation to persistent clusters and inherent structure landscapes, respectively. 
 Coslovich and Pastore used Voronoi tessellation to identify locally preferred structures (LPS), which are connected to the slow dynamics, and showed that their cluster size grows with decreasing temperature \cite{Coslovich2007}. Royall and coworkers used their proposed Topological cluster classification (TCC) \cite{Malins2013JCP} to identify LFS connected to slow particles and studied the growth of these clusters \cite{Malins2013JCP,Dunleavy2012,MalinsLifetimes2013,Royall2014,Crowther2015,Hallett2018,royall2008}. Karmakar \textit{et al.} \cite{KARMAKAR20121001} showed that the minimum eigenvalue of the hessian matrix $\lambda_{min}$ plays a crucial role in defining and understanding the static length at low temperature where an-harmonic effect is absent. Chakrabarty \textit{et al.} \cite{Chakrabarty2017} introduced a new structure agnostic methodology called block analysis to study the static and dynamic lengthscales. 
 The PTS correlation length has also been obtained for many systems \cite{Hocky2012,Berthier2016,russo_tanaka}.
Nevertheless, all these studies show that in systems without MRCO, these structural lengthscales grow more slowly than their dynamical counterparts. Even studies exploring deep supercooled regime below the MCT transition temperature, where the activated dynamics is relevant, have shown that the dynamical lengthscale grows faster than the static PTS lengthscale \cite{berthier_PRX_2022}.

 In the KALJ binary 80:20 mixture of A and B particles, Fernandez and Horowell have shown that although the lattice energy of the total system is lowest when the system crystalizes into CsCl structure between the A and the B particles and the rest of the A particles form face-centered cubic (fcc) structure, isolated CsCl structure was found to be less stable compared to isolated square antiprism structure with B particles at the center \cite{FernandezHarrowell_PRE_2003}. The latter structure is related to the $Al_{2}Cu$ crystal. The presence of this structure was observed by Crowther {\it et al.} using their TCC method \cite{Malins2013JCP} and referred to as 11A structure \cite{Crowther2015}. This 11A structure was found to be an LFS associated with slow dynamics \cite{MalinsLifetimes2013,Royall2014}, but due to compositional frustration, its spatial extent was found not to grow \cite{Crowther2015}, which led to the decoupling of the growth of the static and the dynamical lengthscale~\cite{MalinsLifetimes2013, Royall2014}. More recent investigations suggest that in the KALJ model, the particles do not have a CsCl formation tendency between the two species, but all the A-type particles show a tendency to form a distorted fcc structure where the distortion comes due to the presence of the B particles in between \cite{Ujjwal_jcp_2016}. However, these motifs are not easily captured by standard bond-orientational order or TCC-based descriptors. This disconnect raises an important question: Does the lack of observed static lengthscale growth stem from a true absence of underlying order, or does it reflect the limitations of existing local structural metrics?

In this work, we revisit the Kob Andersen binary Lennard-Jones mixture, examining both the canonical 80:20 composition and a modified 60:40 variant. We begin by analysing the structural lengthscale using a recently developed structural order parameter (SOP) that has been shown to correlate well with particle dynamics~\cite{Manoj_prl_2021,Sharma2022}. However, we find that it does not show a significant growth in spatial correlations upon supercooling and thus fails to reveal a growing static lengthscale. It is important to note that dynamical correlation lengths are typically extracted from long time and not short time, dynamics. Whereas for most order parameters, the success is decided by correlating it with the short time dynamics \cite{liu_nature,Sharma2022,Zhang2020}. Recently, Tong and Tanaka have proposed that for systems with and without MRCO, when the relevant order parameter is coarse grained in space, it provides better correlation with the long time dynamics \cite{TongTanaka2018}. According to their study, the optimal coarse-graining length, at which the structure-dynamics correlation is maximised, increases with decreasing temperature, and this growth is similar to the growth of the dynamical lengthscale ~\cite{TongTanaka2018,TongTanaka2019,TongTanaka2020,Tanaka2025}. There are also other studies which highlight the role of medium range static order in describing the dynamics. A study involving elastically collective nonlinear Langevin equation theory has shown that medium-range order is important in understanding structure dynamics correlation, and the static length then grows with density \cite{ken_sch}.  Similarly, investigations using local volume~\cite{MeiWang2022}, local caging potential~\cite{mohit_wca}, and order parameters derived via unsupervised machine learning~\cite{QiuArun2025} have shown that meaningful correlations with long time dynamics emerge only after spatial coarse-graining of the static order parameter. Here, we demonstrate that the SOP, when coarse-grained over an optimal lengthscale, exhibits strong correlations with long time dynamics. Notably, the static lengthscale extracted from this coarse-grained SOP grows with decreasing temperature in a manner consistent with the dynamical lengthscale. While medium range
structural correlations in MRCO free models have previously been detected in additive binary mixtures \cite{TongTanaka2018,TongTanaka2019,TongTanaka2020,Tanaka2025}, our study successfully identifies such correlations in one of the most
standard glass forming models, the non-additive Kob Andersen model.
The result emphasizes the previous findings \cite{TongTanaka2018,TongTanaka2019,TongTanaka2020,Tanaka2025,ken_sch,MeiWang2022,QiuArun2025} that medium range structural correlations and not purely local order, are essential in understanding the dynamical slowdown in glass-forming systems.

The organisation of the rest of the paper is as follows. Section  \ref{sec:simu} provides the simulation details. In Section  \ref{sec:Analysis}, we describe the methodology used in this work. Section \ref{sec:results} presents the results. Section  \ref{sec:conclusion} presents a concluding discussion. Five appendices are included for additional details.

\section{Simulation Details}
\label{sec:simu}

We simulate the standard 80:20 Kob-Andersen binary Lennard-Jones (KALJ) mixture~\cite{Kob1994} in three dimensions, consisting of particles of types A and B. The interaction between any pair of particles of types \( \alpha \) and \( \beta \), separated by a distance \( r \), is given by 
\begin{equation}
U_{\alpha\beta}(r) = 4 \epsilon_{\alpha\beta} \left[ \left( \frac{\sigma_{\alpha\beta}}{r} \right)^{12} - \left( \frac{\sigma_{\alpha\beta}}{r} \right)^{6} \right]
\end{equation}
where \( \alpha, \beta \in \{ \text{A}, \text{B} \} \) denote the types of the interacting particles. The interaction parameters are set as follows: \( \epsilon_{\mathrm{AA}} = 1.0 \), \( \epsilon_{\mathrm{AB}} = 1.5 \), \( \epsilon_{\mathrm{BB}} = 0.5 \), \( \sigma_{\mathrm{AA}} = 1.0 \), \( \sigma_{\mathrm{AB}} = 0.8 \), and \( \sigma_{\mathrm{BB}} = 0.88 \). All quantities are expressed in reduced LJ units. The number density is fixed at \( \rho = 1.2 \) for the 80:20 mixture and \( \rho = 1.41 \) for the 60:40 mixture. Interactions are truncated and quadratically shifted at a cutoff distance of \( r_c = 2.5\sigma_{\mathrm{AB}} \) to ensure continuity of both the potential and its first derivative~\cite{Stoddard1973}. Simulations are performed in the NVT ensemble using a Nosé-Hoover thermostat with a time step \( \Delta t = 0.005 \). To minimize finite-size effects, we set the system size to \( N = 16{,}000 \) for structural quantity calculations and \( N = 500{,}000 \) for evaluating the four-point correlation function, following the approach outlined in Reference~\cite{Karmakar2010}.

\section{Methodology}
\label{sec:Analysis}

\subsection{Overlap function and four-point structure factor}
\label{sec:s4qt}
The time-dependent overlap function is defined by,
\begin{equation}
\begin{aligned}
Q(t) = \frac{1}{N} \sum_{i=1}^{N} \omega(|\mathbf{r}_i(t) - \mathbf{r}_i(0)|)
\end{aligned}
\label{overlap}
\end{equation}
\noindent
where the function $\omega(x)$ takes the value 1 if $0 \leq x \leq a$ and 0 otherwise and $\mathbf{r}_i(t)$ is the position of particle $i$ at time $t$. \(a = 0.3\sigma\) is typically chosen to account for small particle shifts due to low-amplitude vibrations, which are treated as negligible. This value is comparable to the mean squared displacement (MSD) observed in the plateau region between the ballistic and diffusive regimes.
The \(\alpha\)-relaxation time, \(\tau_\alpha\), is defined as the time at which \(Q(t)\) decays to \(1/e\), i.e., \(Q(\tau_\alpha)=1/e\).

The four-point structure factor \(S_4({\bf{q}},t)\), which characterises the spatial correlations of the dynamical fluctuations in the wavevector plane {\bf{q}}, is given by \cite{Karmakar2010}
\begin{equation}
S_4(\mathbf q,t)=\frac{1}{N}
          \Bigl[\langle Q(\mathbf q,t)Q(-\mathbf q,t)\rangle
          -\langle Q(\mathbf q,t)\rangle^{2}\Bigr]
\label{eq_s4q}
\end{equation}
\noindent
Here $Q(\mathbf q,t)$ is the Fourier Transform of the overlap function given in Eq.\ref{overlap} and can be written as,
\begin{equation}
Q(\mathbf q,t)=\sum_{i=1}^{N} \omega(|\mathbf{r}_i(t) - \mathbf{r}_i(0)|)
               e^{i\mathbf q\cdot\mathbf r_i(0)}
\label{eq:overlap_ft}
\end{equation}
\noindent

To obtain the dynamic correlation length we fit the $S_4(\mathbf q,t)$ at $t=\tau_\alpha$ to an Ornstein–Zernike form in the small $q$ limit as,
\begin{equation}
S_4(\mathbf q,\tau_\alpha)=\frac{S_4( q\!\to\!0,\tau_\alpha)}
               {1+(q\xi_{4})^{2}}
\label{eq_ozs4q}
\end{equation}
where \(\xi_4\) is the dynamic correlation length that grows on approaching the glass transition~\cite{Karmakar2010,Paula2023}. For data analysis, we plot \(1/S_4( q,\tau_\alpha)\) versus $q^{2}$; from Eq.\ref{eq_ozs4q} this should be linear, so an ordinary least-squares fit reliably extracts both the intercept \(1/S_4( q\!\to\!0,\tau_\alpha)\) and the slope \(\xi_4^{2}/S_4( q\!\to\!0,\tau_\alpha)\), which can be then used to obtain $\xi_{4}$.  For this analysis, we use the 500,000 particle system, which allows us to explore smaller q ranges. However, in this small q range, for better averaging, we use a long run length of 200,000 time steps. We fit the data in the range $q=0.1-0.6$.

\subsection{Displacement–displacement correlation function}
\label{sec:guu}

To characterise the spatial extent of correlated motion in supercooled liquids, we use the excess displacement–displacement correlation function $g_{uu}(r,\tau_{\alpha}) = \bigl\langle u(0)\,u(r) \bigr\rangle $ which is defined as, \cite{Poole1998PhysicaA,Donati1999PRL, Paula2023, Hallett2018}.

\begin{equation}
\bigl\langle u(0)\,u(r) \bigr\rangle
= \frac{\Bigl\langle
   \displaystyle\sum_{i,j}
   u(\tau_\alpha)_{i}\,u(\tau_\alpha)_{j}\,
   \delta\!\bigl(r-\lvert\mathbf r_{ij}\rvert\bigr)
   \Bigr\rangle}
  {4\pi r^{2}\,\Delta r\,\rho\,Ng(r)}
\label{eq-gamma_static_1}
\end{equation}

\noindent
where $u_i(\tau_\alpha) = |\mathbf{r}_i(t + \tau_\alpha) - \mathbf{r}_i(t)|$ is the scalar displacement of particle $i$ over time $\tau_\alpha$. g(r) is the radial distribution function (rdf) and is defined as
\begin{equation}
g(r) = 
\frac{
\left\langle 
\sum_{\substack{i,j=1 \\ i \ne j}}^{N} 
\delta\left(r - |\mathbf{r}_{ij}|\right)
\right\rangle
}{
4\pi r^2 \Delta r \rho N
}
\label{eq:gr_correct_form}
\end{equation}

\noindent
where $\rho$ is the number density and N is the number of particles. The normalised excess displacement-displacement correlation function is given by \cite{LandauSchuttler1999}

\begin{equation}
\Gamma_{uu}(r,\tau_{\alpha}) = 
\frac{\bigl\langle u(0)\,u(r)\bigr\rangle
      - \bigl\langle u(\tau_\alpha) \bigr\rangle^{2}}
     {\bigl\langle \bigl(u(\tau_\alpha)\bigr)^{2} \bigr\rangle
      - \bigl\langle u(\tau_\alpha) \bigr\rangle^{2}}
\label{eq_gamma}
\end{equation}

\noindent
 An increasing $\Gamma_{uu}(r,\tau_{\alpha})$ with decreasing temperature indicates growing dynamical heterogeneity near the glass transition.
Assuming that $\Gamma_{uu}(r, \tau_{\alpha})$ decays exponentially with distance, then the associated lengthscale can be extracted by integrating $\Gamma_{uu}(r, \tau_{\alpha})$ over space as shown in the equation below\cite{Poole1998PhysicaA},

\begin{equation}
\xi_{D} = \int_0^\infty \Gamma_{uu}(r, \tau_\alpha) \, dr
\label{ddcr-integration}
\end{equation}
\noindent
yielding $\xi_D$, the characteristic dynamical correlation length. 
For this calculation, we take a 16,000 particle system. The averaging is done over a run length of 950,000 timesteps for 4 configurations.


\subsection{Mean field caging potential as structural order parameter}
\label{sec:SOP}

To characterise the static structure of glass forming systems, we introduce the structural order parameter (SOP)\cite{Manoj_prl_2021,Sharma2022}. The SOP quantifies the extent to which particles are confined by the mean field caging potential resulting from interactions with neighbouring particles, based on the Ramakrishnan–Yussouff free-energy functional~\cite{RamakrishnanYussouff1979}. Here, we distinguish between the macroscopic SOP $\beta\Phi$ and the microscopic SOP $\beta\phi$ by this notation.

The macroscopic effective caging potential at zero displacement is computed as:
\begin{equation}
\beta \Phi = -\rho \int \textbf{dr} \sum_{u,v} C_{uv}(r)\, x_u x_v\, g_{uv}(r),
\label{eq:macro_sop}
\end{equation}
where $\rho$ represents the number density, and $x_u$, $x_v$ indicate mole fractions of particle species $u$ and $v$.
The partial radial distribution function, $g_{uv}(r)$, is given by\cite{HansenMcDonald2013},
\begin{equation}
g_{uv}(r) = \frac{1}{4\pi r^2 \rho_v N_u} \left\langle \sum_{i=1}^{N_u} \sum_{j=1}^{N_v} \delta(r - |\mathbf{r}_i^u - \mathbf{r}_j^v|) \right\rangle
\label{eq:bulk_gr}
\end{equation}
where $\rho_v$ denotes the species-specific number density, $N_u$ and $N_v$ represent the total numbers of particles for species $u$ and $v$, respectively, and the angular brackets represent an ensemble average.

The direct correlation function $C_{uv}(r)$ in this study is approximated using the Hypernetted Chain (HNC) closure relation~\cite{HansenMcDonald2013}:
\begin{equation}
C_{uv}(r) = -\beta U_{uv}(r) + (g_{uv}(r)-1) - \ln g_{uv}(r)
\label{eq:hnc}
\end{equation}
where $U_{uv}(r)$ denotes the interaction potential between particles of species $u$ and $v$, and $\beta = 1/(k_B T)$ is the inverse thermal energy.

At the microscopic scale, the effective caging potential experienced by particle $i$ of species $u$ is given as:
\begin{equation}
\beta \phi^i_{u} = -\rho \int \textbf{dr} \sum_{v} x_v C_{uv}^i(r)\, g_{uv}^i(r)
\label{eq:micro_sop}
\end{equation}
where $\rho$ is the number density, $C_{uv}^i(r)$ is the particle-specific direct correlation function, and $g_{uv}^i(r)$ represents the local radial distribution function around particle $i$.

To obtain $g_{uv}^i(r)$, a Gaussian smoothing method is applied~\cite{PiaggiValssonParrinello2017}:
\begin{equation}
g_{uv}^i(r) = \frac{1}{4\pi \rho r^2} \sum_j \frac{1}{\sqrt{2\pi \delta^2}} \exp\left(-\frac{(r - r_{ij})^2}{2\delta^2}\right)
\label{eq:micro_gr}
\end{equation}
where $r_{ij}$ is the separation between particles $i$ and $j$, and $\delta = 0.09\sigma_{AA}$ is the width of the Gaussian smoothing kernel. This approach yields a smooth, noise-reduced local radial distribution function compared to discrete binning. The correlation between structure and dynamics is best captured when the upper limit of the $r$-integration is set at the position of the first minimum of $g_{uv}(r)$, computed using Eq.~\ref{eq:bulk_gr}. This choice ensures that only the first-nearest neighbours contribute significantly to the SOP. In addition, we impose a lower cutoff on the $r$-integration. This lower bound is selected as the distance at which the average radial distribution function first attains a non-zero value\cite{Sharma2022}.

To avoid unphysical artefacts at short distances in microscopic calculations, we consistently use an approximate expression for the direct correlation function as previously detailed in references~\cite{mohit_wca,PatelSharmaBhattacharyya2023}. Specifically, the simplified form adopted is given by 
\begin{equation}
C_{uv}^{\text{approx}}(r) = g_{uv}(r)-1
\end{equation}
which maintains computational consistency and physical realism throughout the analysis.

Note that while writing the macroscopic caging potential, Eq.\ref{eq:macro_sop} and its microscopic counterpart, Eq.\ref{eq:micro_sop}, we have made a mean field approximation. This implies that we have assumed the background of the tagged particle to be frozen, ignoring its dynamics. Even under this mean field approximation using local rdf, given by Eq.\ref{eq:micro_gr}, it is possible to capture the heterogeneity in the local structure of each particle and thus the distribution of the SOP as shown in Appendix I. We find that with decreases in temperature, the environment around each particle becomes more structured, and the distribution shifts to deeper caging potentials.


\subsection{Structural lengthscale calculation}
\label{sec:beta_phi_corr}

To quantify the structural lengthscale, we use the microscopic caging potential described before as the structural order parameter,
\(\beta\phi\) and follow the methodology used for obtaining the dynamical lengthscale from the displacement-displacement correlation function.
The excess correlation function of the structural order parameter is defined as follows 

\begin{equation}
\bigl\langle \beta\phi(0)\,\beta\phi(r) \bigr\rangle
= \frac{\Bigl\langle
   \displaystyle\sum_{i,j}
   \beta\phi_{i}\,\beta\phi_{j}\,
   \delta\!\bigl(r-\lvert\mathbf r_{ij}\rvert\bigr)
   \Bigr\rangle}
  {4\pi r^{2}\,\Delta r\,\rho\,Ng(r)}
  \label{eq-gamma_static_1}
\end{equation}

The corresponding normalized correlation function is given by, \cite{Cubuk2017,LandauSchuttler1999}
\begin{equation}
\Gamma_{bare}(r) = 
\frac{\bigl\langle \beta\phi(0)\,\beta\phi(r)\bigr\rangle
      - \bigl\langle \beta\phi \bigr\rangle^{2}}
     {\bigl\langle (\beta\phi)^{2} \bigr\rangle
      - \bigl\langle \beta\phi \bigr\rangle^{2}}
\label{eq-gamma_static_bare}
\end{equation}
\noindent
 Integrating $\Gamma_{bare}(r)$ over $r$ yields the characteristic structural lengthscale $\xi_{bare}$,

\begin{equation}
\xi_{bare} = \int_0^\infty \Gamma_{bare}(r) \, dr
\label{bare-integration}
\end{equation}

\noindent
 which is analogous to the procedure used to obtain the dynamical lengthscale (Eq.\ref{ddcr-integration}).

For this calculation, we take 16,000 particle system. The averaging is done over a run length of 200,000 time steps for 12 configurations.


\subsection{Isoconfigurational ensemble}
\label{isoconf}

The iso-configurational ensemble, originally introduced by Harrowell \textit{et al.} \cite{harrowell}, consists of multiple trajectories initiated from the same initial particle configuration but with randomised initial momenta drawn from the Maxwell-Boltzmann distribution at the given temperature. This approach eliminates the trivial variations in particle displacements that result from different initial momenta. Any significant differences observed in the trajectory-averaged dynamics can, therefore, be attributed to the properties of the initial configuration rather than to stochastic variations.

We employ the iso-configurational ensemble method to evaluate the correlation between structure and dynamics. Specifically, we generate 12 independent initial configurations, ensuring they are separated by at least 75$\tau_{\alpha}$ to guarantee their distinction. From each configuration, we initiate 100 independent trajectories. To analyse the resulting dynamics, we define mobility as:

\begin{equation}
\mu^{j}(t) = \frac{1}{N_{IC}}\sum_{i=1}^{N_{IC}} \sqrt{(r_{i}^{j}(t) - r_{i}^{j}(0))^{2}}
\label{eq:mobility}
\end{equation}

Here, $\mu^{j}(t)$ represents the mobility of the $j^{th}$ particle at time $\tau_\alpha$, while $N_{IC}$ denotes the number of independent trajectories. We then quantify the relationship between the initial structure and the dynamics over time using the Spearman rank correlation.

\subsection{Spearman rank correlation}

Spearman rank correlation, a non-parametric measure of statistical dependence, is computed for a dataset containing $m$ values as follows:

\begin{equation}
C_{R}(X,Y) = 1 - \frac{6 \sum d_{i}^2}{m(m^2 -1)}
\label{rank_corrl}
\end{equation}

\noindent
where the squared rank difference is defined as $d_{i}^2 = R(X_{i}) - R(Y_{i})$, with $R(X_{i})$ and $R(Y_{i})$ representing the respective ranks of the raw data points $X_{i}$ and $Y_{i}$.

\section{Results}
\label{sec:results}

\begin{figure}
\centering
\begin{subfigure}{0.4\textwidth}
\includegraphics[width=0.95\linewidth]{Fig1a.eps}
\end{subfigure}
\vskip\baselineskip
\begin{subfigure}{0.4\textwidth}
\includegraphics[width=0.95\linewidth]{Fig1b.eps}
\end{subfigure}
\caption{\textbf{(a)} Inverse of the four-point structure factor, \(1/S_4(\mathbf{q}, \tau_\alpha)\), (Eq.\ref{eq_s4q}) plotted as a function of the squared wavevector, $q^2$, at various temperatures. Solid lines are Ornstein–Zernike (OZ) fits.  \textbf{(b)} Temperature dependence of three characteristic lengthscales: the dynamical lengthscale, $\xi_{4}$ (Eq.\ref{eq_ozs4q}); the dynamical lengthscale, $\xi_{D}$ (Eq.\ref{ddcr-integration}) for A particles; and the bare static lengthcale, $\xi_{bare}$ (Eq.\ref{bare-integration}) for A particles. The scaling factor for $\xi_{4}$ is 2.21, $\xi_D$ is 0.31, \& $\xi_{bare}$ is 0.28. We also plot the PTS lengthscale reported by Hocky {\it et al.} \cite{Hocky2012} scaled by 1.63, the PTS and finite size scaling lengthscale reported by Chakrabarty {\it et al.} \cite{Chakrabarty2017} scaled by 1.60 and the static lengthscale obtained by Zhang and Kob for the angular power spectra \cite{Zhang2020} scaled by 3.52. With a decrease in temperature, both dynamical lengthscales exhibit similar growth, while the static lengthscales show weaker growth.}
\label{fig:primary}
\end{figure}

\begin{figure}
\centering
\begin{subfigure}{0.4\textwidth}
\includegraphics[width=0.95\linewidth]{Fig2a.eps}
\end{subfigure}
\vskip\baselineskip
\begin{subfigure}{0.4\textwidth}
\includegraphics[width=0.95\linewidth]{Fig2b.eps}
\end{subfigure}
\caption{Spearman rank correlation, $C_{R}(\overline{\beta\phi}, \mu)$, between the coarse-grained SOP, $\overline{\beta\phi}$, and mobility, \(\mu\), calculated at times scaled by the \(\alpha\)-relaxation time, \(t/\tau_{\alpha}\), for A-type particles at different coarse-graining lengths, \(L\). \textbf{(a)} Result at high temperature, \(T = 0.70\) and \textbf{(b)} at low temperature, \(T = 0.47\). The colour coding in (b) matches that of (a).}
\label{fig:Cr-vs-time}
\end{figure}


\begin{figure}
\centering
\begin{subfigure}{0.4\textwidth}
\includegraphics[width=0.95\linewidth]{Fig3a.eps}
\end{subfigure}
\vskip\baselineskip
\begin{subfigure}{0.4\textwidth}
\includegraphics[width=0.95\linewidth]{Fig3b.eps}
\end{subfigure}
\vskip\baselineskip
\begin{subfigure}{0.39\textwidth}
\includegraphics[width=0.95\linewidth]{Fig3c.eps}
\end{subfigure}
\caption{Spearman rank correlation, \(C_R(\overline{\beta\phi}, \mu(\tau_{\alpha}))\), between the coarse-grained SOP, $\overline{\beta\phi}$, and mobility, $\mu(\tau_{\alpha})$, calculated at $\alpha$-relaxation time as a function of coarse-graining length, $L$, \textbf{(a)} for A-type particles and \textbf{(b)} for B-type particles at different temperatures. 
The colour coding in (b) matches that of (a)
\textbf{(c)} Temperature dependence of the average value of $L_{max}$, shows stronger growth for A particles and weaker growth for B particles. The error bars show the variation of the $L_{max}$ with configuration. Lines are a guide to the eye.}
\label{fig:Str_dyn82}
\end{figure}

\begin{figure*}
\centering
\begin{subfigure}{0.31\textwidth}
\caption{}
\includegraphics[width=0.98\linewidth]{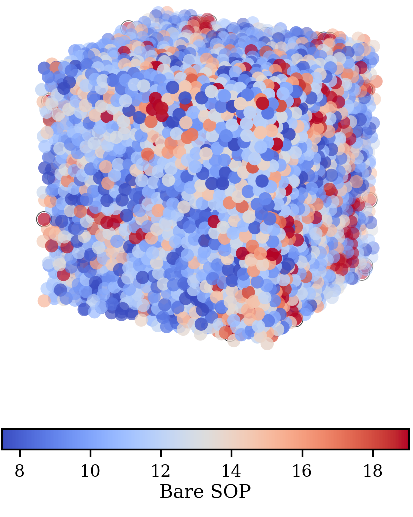}
\end{subfigure}
\begin{subfigure}{0.31\textwidth}
\caption{}
\includegraphics[width=0.98\linewidth]{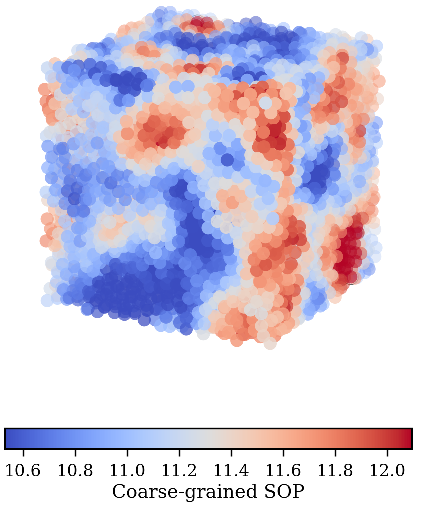}
\end{subfigure}
\begin{subfigure}{0.31\textwidth}
\caption{}
\includegraphics[width=0.98\linewidth]{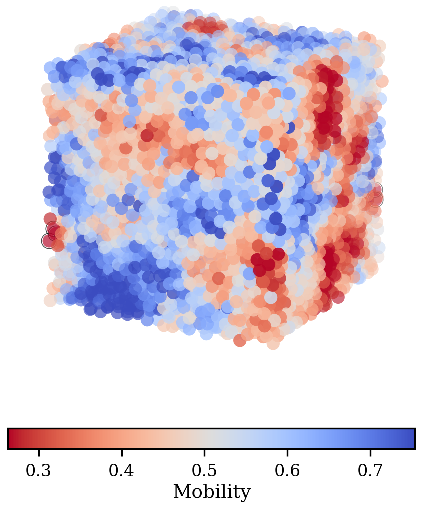}
\end{subfigure}
\begin{subfigure}{0.31\textwidth}
\caption{}
\includegraphics[width=0.98\linewidth]{Fig4d.eps}
\end{subfigure}
\begin{subfigure}{0.31\textwidth}
\caption{}
\includegraphics[width=0.98\linewidth]{Fig4e.eps}
\end{subfigure}
\begin{subfigure}{0.31\textwidth}
\caption{}
\includegraphics[width=0.98\linewidth]{Fig4f.eps}
\end{subfigure}
\caption{Results for 80:20 KALJ system: \textbf{Top panels:} Snapshots at $T=0.45$ showing structural and dynamical fields for A-type particles: \textbf{(a)} The bare SOP, $\beta\phi$, appears spatially noisy, lacking visible large-scale structure. \textbf{(b)} Coarse-graining the SOP, $\overline{\beta \phi}$, with $L_{max}$, shows the emergence of patches of high and low SOP values. \textbf{(c)} The particle mobility field, $\mu (\tau_{\alpha})$, calculated using Eq.\ref{eq:mobility} over the $\alpha$-relaxation time shows low mobility domains that spatially mostly overlap with high SOP regions in (b), indicating a structure–dynamics link. \textbf{Bottom Panels:} Normalized excess correlation functions corresponding to the fields in top panels at different temperatures. \textbf{(d)} The normalised excess correlation of the bare SOP, $\Gamma_{\text{bare}}(r)$, (Eq.\ref{eq-gamma_static_bare}) confirms the absence of significant growth in bare static lengthscale. \textbf{(e)} The normalised excess correlation of the coarse-grained SOP, $\Gamma_{\text{CG}}(r)$, (Eq.\ref{eq-gamma_static_cg}) shows an increasing correlation length with decreasing temperature, consistent with the emergence of extended structural patches. \textbf{(f)} The normalized excess displacement–displacement correlation function, averaged over the configurations $\Gamma_{uu}(r,\tau_\alpha)$, (Eq.\ref{eq_gamma}) exhibits growth with decreasing temperature.
The colour coding in (e) and (f) matches that of (d)}
\label{fig:8020-correlation}
\end{figure*}

\begin{figure}
\centering
\begin{subfigure}{0.5\textwidth}
\includegraphics[width=0.98\linewidth]{Fig5.eps}
\end{subfigure}
\caption{Temperature dependence of the dynamical lengthscale, $\xi_{D}$, and coarse-grained structural lengthscale, $\xi_{CG}$, in the 80:20 KALJ system. $\xi_{D}$ is obtained from  \(\Gamma_{uu}(r,\tau_\alpha)\), (Eq.\ref{ddcr-integration}); and $\xi_{CG}$ obtained from \(\Gamma_{CG}(r)\), (Eq.\ref{CG-integration}). For comparison, the lengthscales are all scaled. The scaling factor for $\xi_{D}$ A-type is 0.31, $\xi_{D}$ B-type is 0.22, $\xi_{CG}$ A-type is 1.00, \& $\xi_{CG}$ B-type is 6.489. The dotted line is a power law fit with $\xi=\xi_{o} [(T-T_{VFT})/T_{VFT}]^{-2/3}$, where $T_{VFT}=0.319$, same as that obtained for the relaxation time. Inset: We plot $ln(\tau_{\alpha})$ against $\xi^{3/2}$ for $\xi_{4}$, $\xi_{D}$ and $\xi_{CG}$ for A particles. The linear fit confirms a power law relationship between the dynamics and the lengthscales.}
\label{fig:8020_compa_lengthscale}
\end{figure}

\begin{figure}
\centering
\begin{subfigure}{0.47\textwidth}
\includegraphics[width=0.98\linewidth]{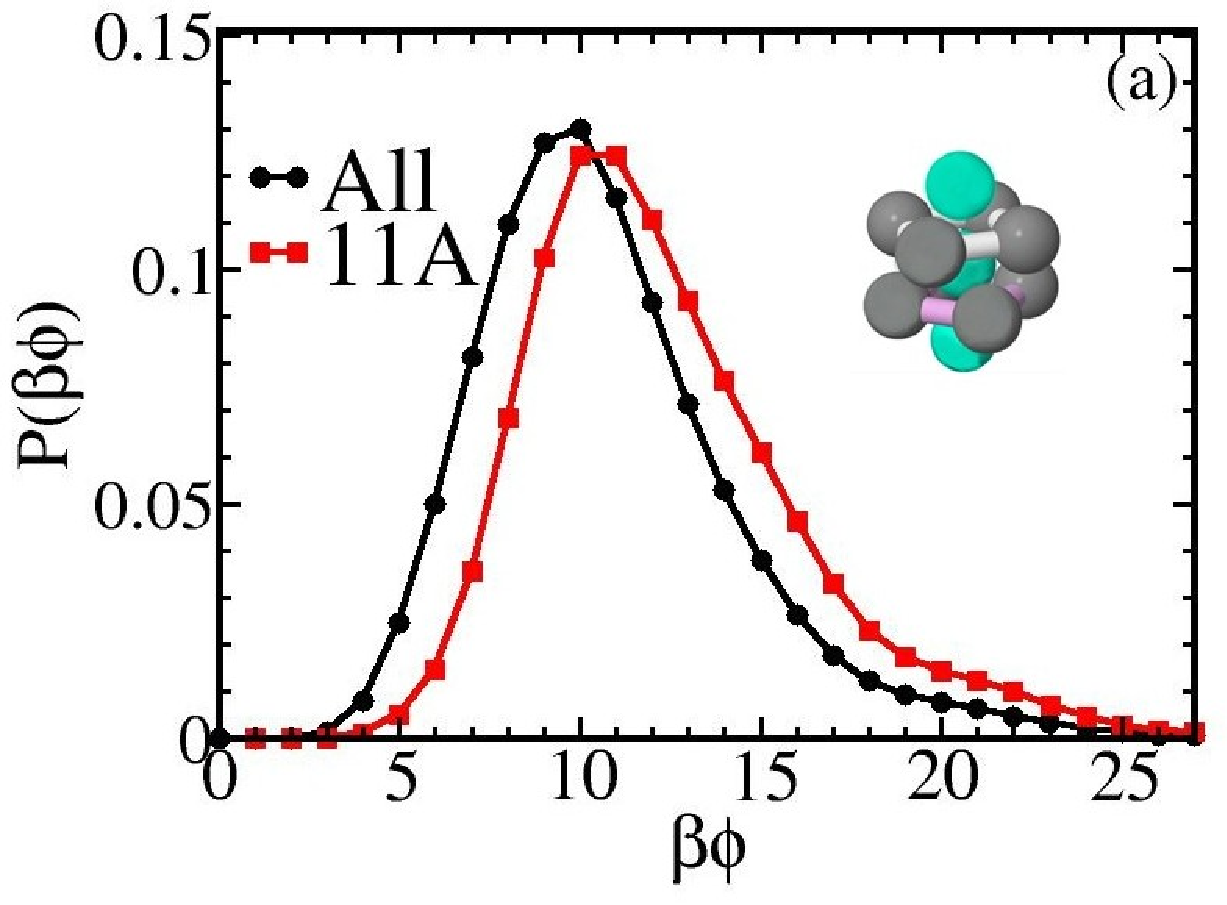}
\end{subfigure}
\vskip\baselineskip
\begin{subfigure}{0.45\textwidth}
\includegraphics[width=0.98\linewidth]{Fig6b.eps}
\end{subfigure}
\caption{\textbf{(a)} Distribution of the bare SOP, $\beta \phi$, for all particles (black, circle) and for those belonging to 11A TCC clusters (red, square). The 11A particles exhibit a pronounced shift toward deeper cages. The 11A structure, which is a bicapped square antiprism, is also shown. \textbf{(b)} Overlap function (Eq.\ref{overlap}) for all particles (black, circle), 11A particles (red, square), and the 2000 most deeply caged particles (top 12.5\%) (green, triangle). Both 11A and 2000 deeply caged particles show significantly slower dynamics than the average.}
\label{fig:tcc_overlap}
\end{figure}


\begin{figure}
\centering
\begin{subfigure}{0.4\textwidth}
\includegraphics[width=0.95\linewidth]{Fig7a.eps}
\end{subfigure}
\vskip\baselineskip
\begin{subfigure}{0.4\textwidth}
\includegraphics[width=0.95\linewidth]{Fig7b.eps}
\end{subfigure}
\caption{Normalized excess displacement–displacement correlation $\Gamma_{uu}(r,\tau_\alpha)$, (Eq.\ref{eq_gamma}) at different temperatures in the 60:40 mixture. \textbf{(a)} A-type particles.\textbf{(b)} B-type particles. The colour coding in (b) matches that of (a).}
\label{fig:ddcr_type_6040}
\end{figure}


\begin{figure*}
\centering

\begin{subfigure}{0.3\textwidth}
  \begin{overpic}[width=0.98\linewidth]{Fig8a.eps}
    \put(2,92){{(a)}}
  \end{overpic}
\end{subfigure}
\hspace{0.2cm}
\begin{subfigure}{0.3\textwidth}
  \begin{overpic}[width=0.98\linewidth]{Fig8b.eps}
    \put(2,92){{(b)}}
  \end{overpic}
\end{subfigure}
\hspace{0.2cm}
\begin{subfigure}{0.3\textwidth}
  \begin{overpic}[width=0.98\linewidth]{Fig8c.eps}
    \put(2,92){{(c)}}
  \end{overpic}
\end{subfigure}
\vspace{0.4cm}
\begin{subfigure}{0.23\textwidth}
  \begin{overpic}[width=0.98\linewidth]{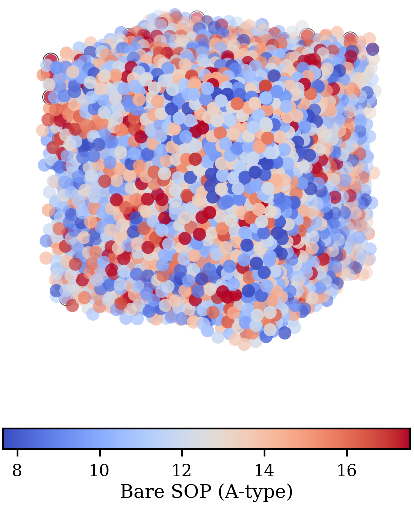}
    \put(2,93){{(d)}}
  \end{overpic}
\end{subfigure}
\begin{subfigure}{0.23\textwidth}
  \begin{overpic}[width=0.98\linewidth]{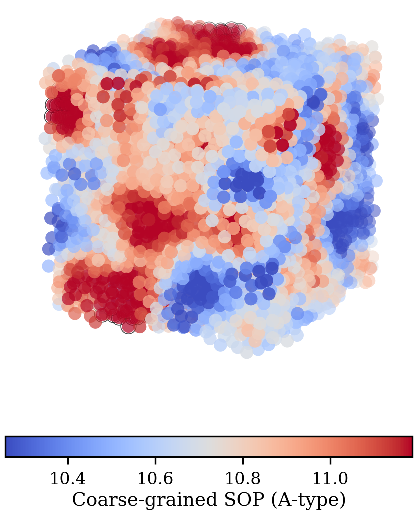}
    \put(2,93){{(e)}}
  \end{overpic}
\end{subfigure}
\begin{subfigure}{0.23\textwidth}
  \begin{overpic}[width=0.98\linewidth]{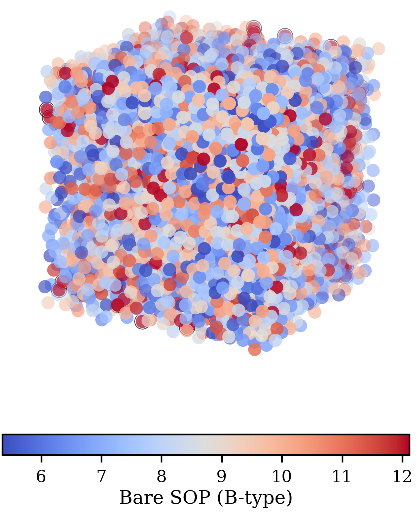}
    \put(2,93){{(f)}}
  \end{overpic}
\end{subfigure}
\begin{subfigure}{0.23\textwidth}
  \begin{overpic}[width=0.98\linewidth]{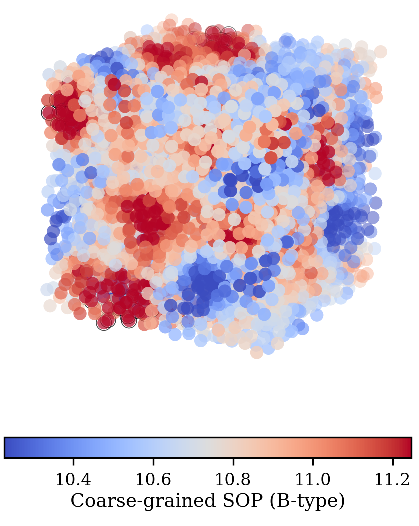}
    \put(2,93){{(g)}}
  \end{overpic}
\end{subfigure}
\vskip\baselineskip
\begin{subfigure}{0.23\textwidth}
  \begin{overpic}[width=0.98\linewidth]{Fig8h.eps}
    \put(2,80){{(h)}}
  \end{overpic}
\end{subfigure}
\begin{subfigure}{0.24\textwidth}
  \begin{overpic}[width=0.98\linewidth]{Fig8i.eps}
    \put(2,80){{(i)}}
  \end{overpic}
\end{subfigure}
\begin{subfigure}{0.23\textwidth}
  \begin{overpic}[width=0.98\linewidth]{Fig8j.eps}
    \put(2,80){{(j)}}
  \end{overpic}
\end{subfigure}
\begin{subfigure}{0.24\textwidth}
  \begin{overpic}[width=0.98\linewidth]{Fig8k.eps}
    \put(2,80){{(k)}}
  \end{overpic}
\end{subfigure}

\caption{Results for 60:40 KALJ system: \textbf{Top panels:} Spearman rank correlation, $C_{R}(\overline{\beta\phi}, \mu(\tau_{\alpha}))$, between the coarse-grained structural order parameter, $\overline{\beta\phi}$, and mobility, $\mu(\tau_{\alpha})$, calculated at the $\alpha$-relaxation time for various coarse-graining lengths $L$, for \textbf{(a)} A-type particles at various temperatures and \textbf{(b)} B-type particles at various temperatures.  \textbf{(c)} Temperature dependence of $L_{max}$, the value of $L$ in (a) and (b) at which $C_R(\overline{\beta\phi}, \mu(\tau_{\alpha})$ peaks, shows strong growth for A particles and B particles. Lines are a guide to the eye.\textbf{Middle panels:} Snapshot at T=0.45 for the structural fields: \textbf{(d)} bare SOP, $\beta\phi$, for A-type particles; \textbf{(e)} coarse-grained SOP, $\overline{\beta\phi}$, for A-type particles; \textbf{(f)} bare SOP, $\beta\phi$, for B-type particles; and \textbf{(g)} coarse-grained SOP, $\overline{\beta\phi}$, for B-type particles. 
Bare SOP appears spatially noisy, lacking visible large scale structure. Coarse-graining the SOP, with $L_{max}$, shows the emergence of patches of high and low SOP values. \textbf{Bottom panels:} Excess correlation functions corresponding to the fields in the middle panels at different temperatures: \textbf{(h)} bare correlation function, $\Gamma_{\text{bare}}(r)$ (Eq.\ref{eq-gamma_static_bare}), for A-type particles; \textbf{(i)} coarse-grained correlation function, averaged over the configurations, $\Gamma_{\text{CG}}(r)$ (Eq.\ref{eq-gamma_static_cg}), for A-type particles; \textbf{(j)} bare correlation function, $\Gamma_{\text{bare}}(r)$ (Eq.\ref{eq-gamma_static_bare}), for B-type particles; and \textbf{(k)} coarse-grained correlation function, averaged over the configurations,  $\Gamma_{\text{CG}}(r)$ (Eq.\ref{eq-gamma_static_cg}), for B-type particles. 
While bare SOP fields show little spatial correlation due to their noisy nature, the coarse-grained fields exhibit clear growth in spatial correlations, consistent with the emergence of structural patches. The colour coding in (b),(h),(i),(j), and (k) matches that of (a).}

\label{fig:6040ranksop}
\end{figure*}

\begin{figure}
\centering
\begin{subfigure}{0.45\textwidth}
\includegraphics[width=0.98\linewidth]{Fig9.eps}
\end{subfigure}
\caption{
Temperature dependence of the dynamical lengthscale, $\xi_{D}$, and coarse-grained structural lengthscale, $\xi_{CG}$, in the 60:40 KALJ system for the A and B types of particles. $\xi_{D}$ is obtained from  \(\Gamma_{uu}(r, \tau_\alpha)\), (Eq.\ref{ddcr-integration}); and $\xi_{CG}$ is obtained from \(\Gamma_{CG}(r)\), (Eq.\ref{CG-integration}). For comparison, the lengthscales are all scaled. The scaling factor for $\xi_{D}$ A-type is 0.30, $\xi_{D}$ B-type is 0.25, $\xi_{CG}$ A-type is 1.24, \& $\xi_{CG}$ B-type is 2.38. The dotted line is a power law fit with $\xi=\xi_{o} [(T-T_{VFT})/T_{VFT}]^{-2/3}$, where $T_{VFT}=0.312$, same as that obtained from the relaxation time. Inset: We plot $ln(\tau_{\alpha})$ against $\xi^{3/2}$ for $\xi_{4}$, $\xi_{D}$ and $\xi_{CG}$ for A and B particles. The linear fit confirms a power law relationship between the dynamics and the lengthscales.}
\label{fig:6040_compa_lengthscale}
\end{figure}

We first investigate the correlation between the dynamical and static lengthscales in the 80:20 Kob Andersen binary Lennard Jones mixture (KALJ). 

There are different formalisms to calculate the dynamical lengthscale. In Fig.\ref{fig:primary}(a) we plot $1/S_{4}(\mathbf{q},\tau_{\alpha})$ (Section \ref{sec:s4qt}) as a function of $q^{2}$ and from this fitting using Ornstein Zernike expression (Eq.\ref{eq_ozs4q}) we obtain the dynamical correlation length $\xi_4$. In Appendix II, we plot the $S_{4}({q},\tau_{\alpha})/S_{4}({q=0},\tau_{\alpha})$ as a function of $q\xi_{4}$ which shows a master plot following the OZ relationship. As we know from previous studies, this data collapse predicts the reliability of the OZ fitting \cite{Szamel_s4qt,Tah_dis_dis}.  
Using the displacement-displacement correlation function, $g_{uu}(r, \tau_{\alpha})$, we calculate the excess displacement-displacement correlation function, $\Gamma_{uu}(r,\tau_{\alpha})$ (Eq.\ref{eq_gamma}) for the A particles, plotted later in Fig.\ref{fig:8020-correlation}(f). Integration of this excess correlation gives us another measure of the dynamic lengthscale, $\xi_{D}$ (Eq.\ref{ddcr-integration}) \cite{Poole1998PhysicaA}.

Here, we study the static lengthscale as described by the recently developed structural order parameter, the depth of the mean field caging potential, $\beta\phi$ ~\cite{Manoj_prl_2021,Sharma2022}. To be consistent with our methodology, similar to the way the dynamical lengthscale is obtained from the normalised excess displacement-displacement correlation, we use the normalised excess correlation of the SOP to compute the static lengthscale as given by Eq.\ref{eq-gamma_static_bare}. The lengthscale $\xi_{bare}$ for the A particles is then obtained from the area under the curve (Eq.\ref{bare-integration}) of the normalised excess correlation of the bare SOP for the A particles shown later in Fig.\ref{fig:8020-correlation}(d). In Fig.\ref{fig:primary}(b), we plot the two dynamical lengthscales and the static lengthscale. We find that both the dynamical lengthscales grow similarly; however, the static lengthscale, grows much slowly. In the same figure for comparison, we also plot the PTS lengthscale reported by Hocky {\it et al.} \cite{Hocky2012}, the PTS and finite size scaling lengthscale reported by Chakrabarty {\it et al.} \cite{Chakrabarty2017}, and the static lengthscale obtained by Zhang and Kob for the angular power spectra \cite{Zhang2020}. We find that in the temperature regime that we have studied, our static lengthscale, the PTS lengthscale, and that associated with the angular power spectra grow in a similar fashion, more slowly than the dynamic lengthscales.

In order to further investigate how the structural and the dynamical lengthscales correlate, we first examine the structure-dynamics correlation.
This correlation is studied using isoconfigurational ensembles. Isoconfigurational ensembles (Section \ref{isoconf}) were designed by Harrowell and coworkers \cite{harrowell} to disentangle the influence of the initial structure from thermal noise in the dynamics.
In this part of the analysis, to obtain cleaner statistics, we separately deal with the A and the B particles in the system.
The results of the structure-dynamics correlation, given by the Spearman rank correlation, $C_R({\beta\phi},\mu)$ between the bare SOP and the mobility for the $A$ particles are given in the main text, and those for the $B$ particles are given in Appendix IV.
As shown in Fig.\ref{fig:Cr-vs-time}(a), $C_R({\beta\phi},\mu)$ between the local per particle level bare SOP with the mobility for the $A$ particles is high at shorter times, but at longer times, there is a drop in the correlation. This drop is stronger at lower temperatures as seen in Fig.\ref{fig:Cr-vs-time}(b). The results are similar for the $B$ particles (Fig.\ref{fig:Cr-vs-time-B} in Appendix IV). This indicates that the local bare SOP has strong predictive power for short time dynamics, but its influence diminishes at longer times, particularly at low temperatures. 

As mentioned in Section \ref{sec:s4qt} and Section \ref{sec:guu}, the dynamical heterogeneity lengthscale is typically evaluated at the time scale, $\tau_\alpha$ where the spatial correlations of particle displacements are most pronounced and dynamic heterogeneity reaches its maximum.~\cite{Berthier_Biroli_PRE_2005}
Thus, any static order parameter that does not show a good correlation at that timescale may not be able to predict the correct static lengthscale. 

Recently, there has been a large number of studies \cite{TongTanaka2018,das_possible_2018,TongTanaka2019,TongTanaka2020,Tanaka2025,MeiWang2022,QiuArun2025}, including us \cite{mohit_wca}, where it is shown that if we coarse-grain the SOP, then the structure-dynamics correlation increases at longer times. The coarse-grained SOP, $\overline{\beta\phi}_i(L)$, is defined as 

\begin{equation}
\overline{\beta\phi}_i(L)
= \frac{
   \displaystyle\sum_{j\neq i}\beta\phi(\mathbf r_j)\,\exp\bigl[-|\mathbf r_i-\mathbf r_j|/L\bigr]
   \;+\;\beta\phi(\mathbf r_i)
  }
  {
   \displaystyle\sum_{j\neq i}\exp\bigl[-|\mathbf r_i-\mathbf r_j|/L\bigr]
   \;+\;1
  }\,
\label{eq:beta_phi_cg_self}
\end{equation}
\noindent
where the sum over \(j\) includes all particles within a distance \(L\) of the tagged particle \(i\).
This coarse-graining reflects the influence of a particle's local environment, weighted exponentially by spatial proximity, and thus captures collective structural effects beyond the single-particle scale. In Fig.\ref{fig:Cr-vs-time}(a), we plot the configuration averaged Spearman rank correlation $C_R(\overline{\beta\phi},\mu)$ against scaled time $t/\tau_{\alpha}$ for different values of coarse-graining length. We find that at short times for higher $L$ values, the correlation is poor, and then at longer times, the correlation increases, and at even longer times, it decreases. We also find that this increase in correlation at longer times initially increases with \(L\) and eventually, for higher \(L\) values, the correlation drops. This implies there is an optimum value of $L$, we refer to as $L_{max}$ where the correlation is maximum. When we compare the plots in Fig.\ref{fig:Cr-vs-time}(a) and \ref{fig:Cr-vs-time}(b), we find that $L_{max}$ increases with decreasing temperature. It is evident from Fig.\ref{fig:Cr-vs-time} that the value of $L_{max}$ also depends on the time at which the mobility is calculated. 

As the dynamical lengthscale is calculated at $\tau_{\alpha}$, we next analyse the correlation and the corresponding value of $L_{max}$ for the mobility at $t=\tau_{\alpha}$. In Fig.\ref{fig:Str_dyn82}(a) and Fig.\ref{fig:Str_dyn82}(b)  we plot the configuration averaged Spearman rank correlation $C_R(\overline{\beta\phi},\mu(\tau_{\alpha}))$, between the coarse-grained SOP and the mobility calculated at $t=\tau_{\alpha}$ for the $A$ and $B$ particles, respectively, against different coarse-graining length, $L$. This plot clearly shows that at each temperature, there is an optimum value of $L$ for which
the correlation is maximum. For the estimation of the value of $L_{max}$ instead of using the configuration averaged Spearman Rank correlation we plot the $C_R(\overline{\beta\phi},\mu(\tau_{\alpha}))$ for every configuration and obtain the peak position which is the corresponding $L_{max}$ value for that configuration. We find that there is a variation of the $L_{max}$ value with configuration. The average of these $L_{max}$ values is plotted in Fig.\ref{fig:Str_dyn82}(c), and the spread of the data is represented by the error bars. Both for A and B particles, this average $L_{max}$ value increases with a decrease in temperature. 
 
However, at low temperatures, the $L_{max}$ value for the $B$ particles grows less than that of the $A$ particles (Fig.\ref{fig:Str_dyn82}(c)), and this difference increases with a decrease in temperature. Note that although we study the A and the B particles separately, while coarse-graining the SOP, if the neighbouring particle is of a different species, its value is included in the coarse-graining process. Thus, the lower value of $L_{max}$ for the $B$ particles is not due to a lower concentration of the $B$ particles. This just shows that, compared to that for the A particles, fewer particles contribute to the long time dynamics of the $B$ particles. 

In Fig.\ref{fig:8020-correlation}(a-c), we show the snapshots of the single particle bare SOP, coarse-grained SOP, and also the mobility at $t=\tau_{\alpha}$ for the A particles at $T=0.45$. We find that the snapshot of the single particle bare SOP (Fig.\ref{fig:8020-correlation}(a)) is highly fluctuating data and has no correlation with the mobility at long times (Fig.\ref{fig:8020-correlation}(c)). However, when we plot SOP, which is coarse-grained over $L_{max}=3.56$ we find regions of high and low values of SOP (Fig.\ref{fig:8020-correlation}(b)) which visually correlates with the regions of high and low values of mobility (Fig.\ref{fig:8020-correlation}(c))  It is this correlation between the coarse-grained SOP and the mobility that is quantified in terms of Spearman rank correlation values given in Fig.\ref{fig:Str_dyn82}(a). 
From the above plots, a key observation is that once we introduce a coarse-graining length $L_{max}$, the ability of SOP to predict which particles will move significantly over long times improves. Physically, this indicates that the long time rearrangements depend not just on the SOP value of a single particle but also on the SOP values of its surrounding particles, giving rise to the medium range order. The lengthscale of this medium range order also increases with a decrease in temperature, a concept that is in tune with the growing cooperatively rearranging regions.

Our study also suggests that the static order parameter itself changes with the timescale over which we measure the dynamics. While studying short time dynamics, the static order parameter that correlates with the dynamics should be calculated at a single particle level, as seen in Fig.\ref{fig:Cr-vs-time}(a) (black circles) and in earlier studies involving some of us \cite{Sharma2022}. Note that a similar observation was also made for the machine learning studies of the softness parameter by Liu and coworkers \cite{liu_nature, Cubuk2017}. However, while studying properties related to long time dynamics like the dynamical correlation length, the corresponding static order parameter is no longer a single particle order parameter but a coarse-grained quantity, as seen in Fig\ref{fig:Cr-vs-time}(b) (pink triangles). This observation is similar to that made in earlier studies \cite{QiuArun2025,TongTanaka2019,MeiWang2022,mohit_wca,ken_sch}.  

Thus, to extract the correct static lengthscale which is commensurate with the dynamical lengthscale we reassign the SOP value of each particle with its coarse-grained value and calculate the normalised excess correlation between the coarse-grained SOP as,

\begin{equation}
\Gamma_{CG}(r)=
\frac{\bigl\langle \overline{\beta\phi}(0)\,\overline{\beta\phi}(r)\bigr\rangle
      - \bigl\langle \overline{\beta\phi} \bigr\rangle^{2}}
     {\bigl\langle (\overline{\beta\phi})^{2} \bigr\rangle
      - \bigl\langle \overline{\beta\phi} \bigr\rangle^{2}}
\label{eq-gamma_static_cg}
\end{equation}
\noindent 
and the corresponding coarse-grained static lengthscale, $\xi_{CG}$ is given by, 
\begin{equation}
\xi_{CG} = \int_0^\infty \Gamma_{CG}(r) \, dr
\label{CG-integration}
\end{equation}

For each iso configurational run $\Gamma_{CG}(r)$ and the corresponding $\xi_{CG}$ is calculated by using the corresponding $L_{max}$ value. In Fig.\ref{fig:8020-correlation}(d), we now plot the  $\Gamma_{bare}(r)$, which is the excess correlation between the bare SOP values of the A particles, which are calculated at the single particle level, and find that as a function of temperature, it grows moderately. However, when we plot the configuration averaged $\Gamma_{CG}(r)$, which is the excess correlation between the coarse-grained SOP values, we find that $\Gamma_{CG}(r)$ for the A particles at lower temperatures survives till longer distances. This clearly shows that the static lengthscale related to the coarse-grained order parameter grows. Note that this growth of excess correlation of the coarse-grained SOP at low temperatures can be visually seen in the form of patches of high and low SOP values in Fig.\ref{fig:8020-correlation}(b). In Fig.\ref{fig:8020-correlation}(f), we also plot the excess displacement-displacement correlation only for the $A$ particles. The plots for the excess correlation for the bare and coarse-grained SOP values and the excess displacement-displacement correlation for the B particles are plotted in Fig.\ref{fig:sop_Btype_8020} in Appendix IV.

 The static lengthscales obtained from $\Gamma_{CG}$ for every configuration are also separately calculated for the A and the B particles by coarse-graining their SOP values till their respective $L_{max}$ values obtained for that configuration. The average of this static lengthscale with error bars is plotted in Fig.\ref{fig:8020_compa_lengthscale} along with the dynamic lengthscales.
We find that the dynamic lengthscale of both species grows similarly. However, the static lengthscale of only the A particles grows like the dynamical lengthscale, but that of the B particles grows moderately, as shown in Fig.\ref{fig:8020_compa_lengthscale}. In Appendix IV in Fig.\ref{fig:xi_vs_Lmax}, plot the $L_{max}$ against the $\xi_{CG}$ of the $A$ particles, and we find that they show a near linear dependence, suggesting that $\xi_{CG}$ grows in tandem with $L_{max}$.

In Fig.\ref{fig:vft_fitting} in Appendix III we show that the relaxation time can be fitted to a Vogel-Fulcher-Tammann (VFT) expression, $\tau_{\alpha}=\tau_{o}*exp(\frac{T_{VFT}}{K_{VFT}(T-T_{VFT})})$ where $\tau_{o}$ is the prefactor, $K_{VFT}$ is the fragility which quantifies how fast the dynamics diverges and  $T_{VFT}$ is the temperature where the dynamics is expected to diverge. Similarly, the temperature dependence of the lengthscale is also given by the following expression $\xi=\xi_{o} [(T-T_{VFT})/T_{VFT}]^{-2/3}$, where $\xi_{o}$ is a prefactor. In Fig.\ref{fig:8020_compa_lengthscale} we show that this functional form can reasonably fit the static and the dynamic lengthscales. This implies that there is a power law dependence between the relaxation time and the correlation length, where $ln(\tau_{\alpha})$ grows as $\xi^{d/2}$, shown earlier in other studies, where `d' is the dimension \cite{  tanakamrco_critical_natmat_2010,tanakamrco_kawasaki_prl_2007,TongTanaka2018}. In the inset of Fig.\ref{fig:8020_compa_lengthscale} we plot $ln(\tau_{\alpha})$ against $\xi^{3/2}$, which can be fitted to a straight line, confirming the relationship. This temperature dependence of the lengthscales and its correlation with the dynamics is consistent with  Ising-type critical scenario \cite{tanakamrco_critical_natmat_2010,Tanaka2012bondliquid, Langer2013}and random first order transition theory in finite dimensions\cite{Lubchenko2007,Kirkpatrick1989}.

Interestingly, many of the earlier studies, trying to find the static correlation length for an 80:20 KALJ system, worked with LFS, like the 11A structure obtained from TCC \cite{Malins2013JCP}, which is also the ⟨0,2,8⟩ Voronoï polyhedra \cite{Coslovich2011}. This structure was found to be long lived, thus contributing to the slow dynamics \cite{Royall2014,Coslovich2011}. Here we plot the distribution of the SOP of particles in the 11A structure obtained via TCC \cite{Malins2013JCP} along with the distribution of the SOP of all the particles in Fig.\ref{fig:tcc_overlap}(a). We find that the SOP values of the 11A particles are shifted towards a higher caging potential, suggesting that these particles are indeed in a deeper caging potential. However, the distribution is also quite wide, which implies that, unlike proper crystals, the local arrangements of the particles in the 11A structure vary over a wide range. We plot the overlap function of the particles that are in the 11A structure, along with the overlap function of all the particles in Fig.\ref{fig:tcc_overlap}(b). We find that, indeed, the particles belonging to the 11A structure are slower compared to the average dynamics.
Note that at $T=0.45$ on average, there are about $12.5\%$ particles in the 11A structure. For a system of 16,000 particles, this comes to about 2000 particles. So we calculate the overlap function for the 2000 particles which have the deepest caging potential and plot it in Fig.\ref{fig:tcc_overlap}(b). We find that they are marginally slower than the 11A particles. Thus, we show that our SOP can also identify slow moving particles, and like the recently observed anticorrelation between the LFS and the excitations, there will be an anticorrelation between the particle in the deepest minima and the excitations \cite{Royall_excitations_2025}.\\

We next discuss the importance of coarse-graining the order parameter to obtain a proper lengthscale. Our results show that a single particle structural property can capture dynamical behaviour only at short timescales. In contrast, the dynamics and associated properties at the $\alpha$ relaxation timescale depend not only on the immediate structural environment of a tagged particle but also on that of its neighbours. For example, a particle in a shallow caging potential may appear mobile at short times, leading to a strong correlation between the structural order parameter and short-time dynamics. However, for the particle to remain mobile over longer times, its neighbours must also reside in shallow cages and thus be more mobile. This collective effect of the surrounding environment is naturally incorporated in a coarse-grained SOP, which, by including both the particle’s structure and that of its neighbours, provides the essential information needed to describe long-time dynamics. It is instructive to compare our findings with the earlier study of Zhang and Kob \cite{Zhang2020}, who correlated a structural order parameter with dynamics at the shorter timescale corresponding to the peak of the non-Gaussian parameter, $t^{*}$. The difference between $t^{*}$ and the structural relaxation time $\tau_{\alpha}$ increases as temperature decreases, with $\tau_{\alpha}$ becoming much larger. Consequently, their order parameter, like our bare SOP, showed little growth at lower temperatures. Our study suggests that this slow growth arises because an order parameter that successfully predicts short time dynamics is not necessarily appropriate for describing dynamics and dynamical properties at longer times.

We next study the growth of the static and dynamic lengthscale in a 60:40 mixture. This system is studied at a higher density $\rho=1.41$ so that the temperature range is similar to that of the 80:20 mixture (Fig.\ref{fig:vft_fitting} in Appendix III). In Fig.\ref{fig:ddcr_type_6040}(a) and Fig.\ref{fig:ddcr_type_6040}(b), we plot the excess displacement-displacement correlation functions of the A and the B particles, respectively. We find that, similar to the 80:20 mixture, both for the A and B particles, the excess correlation grows with the lowering of the temperature. In Fig.\ref{fig:6040ranksop}(a) and Fig.\ref{fig:6040ranksop}(b), we plot the Spearman rank correlation of the coarse-grained SOP and the mobility at $t=\tau_{\alpha}$ for the A and B particles, respectively. Both show that there is an optimum value of coarse-graining length $L_{max}$ where the correlation is maximum, and $L_{max}$ grows with a decrease in temperature. Unlike in the case for the 80:20 mixture, in this case (Fig.\ref{fig:6040ranksop}(c)), the $L_{max}$ for the $B$ particles grows almost as much as that for the $A$ particles.

In Fig.\ref{fig:6040ranksop}(d) and Fig.\ref{fig:6040ranksop}(f), we show the snapshots of the bare SOP values for the A and the B particles, respectively, at $T=0.45$. These show that the values of the SOP fluctuate at the single particle level. In Fig.\ref{fig:6040ranksop}(e) and Fig.\ref{fig:6040ranksop}(g), we show snapshots of the coarse-grained SOP for the A and the B particles, respectively. We find that the coarse-grained SOP does show patches of regions with low and high SOP values. In Fig.\ref{fig:6040ranksop}(h)
and Fig.\ref{fig:6040ranksop}(j), we plot $\Gamma_{bare}$ (Eq.\ref{eq-gamma_static_bare}) at different temperatures, and they show a weak temperature dependence. However, $\Gamma_{CG}$ (Eq.\ref{eq-gamma_static_cg}) for the A and the B particles show that with a decrease in temperature, the correlation survives till a longer distance, suggesting a growing correlation length (Fig.\ref{fig:6040ranksop}(i) and Fig.\ref{fig:6040ranksop}(k)). To quantify these results in Fig.\ref{fig:6040_compa_lengthscale}, we plot the dynamic and the static lengthscales for the A and the B particles. We find that in this case, both the static and the dynamical lengthscales for both species grow together. 
Moreover, similar to the 80:20 mixture, we show that even for the 60:40 mixture, the expression $\xi=\xi_{o} [(T-T_{VFT})/T_{VFT}]^{-2/3}$ can fit the static and the dynamic lengthscales reasonably well. In the inset, we also plot $ln(\tau_{\alpha})$ against $\xi^{3/2}$, which confirms the power law relationship between the relaxation time and the lengthscales. 

\section{Conclusions}
\label{sec:conclusion}

In this work, we investigate the relationship between structural and dynamical lengthscales in Kob Andersen binary Lennard Jones (KALJ) mixtures with 80:20 and 60:40 compositions, focusing on how these lengthscales evolve as the systems are supercooled. Using displacement–displacement correlation functions, we compute the dynamic correlation length, $\xi_{D} $, which exhibits clear growth with decreasing temperature, consistent with increasing dynamic heterogeneity near the glass transition. We further find that the dynamical lengthscales for both A and B particles grow in a similar fashion when analysed separately.

To examine the static counterpart, we analyse lengthscales derived from a recently developed structural order parameter (SOP) based on a mean field caging potential \cite{Manoj_prl_2021,Sharma2022}. When using the local bare SOP at a single particle level, the static lengthscale shows limited growth, echoing earlier observations \cite{Hocky2012,KARMAKAR20121001,Biroli2013,Berthier2016,Yaida2016,MalinsLifetimes2013,Royall2014}. This result is initially surprising given previous studies involving some of us demonstrated a strong correlation between this SOP and dynamics \cite{Sharma2022,mohit_wca}. To resolve this, we analyse the structure–dynamics correlation across different timescales. We find that while the local SOP correlates well with short time dynamics, its correlation weakens over long timescales. This is because long time rearrangements are influenced not only by the SOP of a given particle but also by the SOP of its neighbours, indicating the emergence of medium range order. 

By coarse-graining the SOP over a spatial lengthscale, $L$, we identify an optimal coarse-graining length, $L_{max}$, that maximises structure–dynamics correlation at $\tau_\alpha$ time scale. We find that this $L_{max}$ grows with a decrease in temperature. For the 80:20 mixture, this growth is more pronounced for the A particles than for the B particles, while for the 60:40 mixture, both species show comparable growth. The corresponding static correlation length, extracted from the spatial correlations of the coarse-grained SOP, also increases significantly at low temperatures. Moreover, we find that for the 80:20 mixture, only the static lengthscale of A particles grows with cooling, and in the 60:40 mixture, the static lengthscales of both species grow in tandem with the dynamical lengthscale. At this point, we do not completely understand why the structural lengthscale for the B particles does not grow, but as discussed later, there might be some correlation with the underlying crystalline order.

These findings provide important insight into the long standing question of whether structural order underlies the slowing down of dynamics in glass forming liquids. Our results indicate that the structural order parameter itself changes depending on the timescale over which the dynamical quantities are studied. For short time dynamics, the bare structural order parameter is a good order parameter; however, to understand the properties connected to the long time dynamics, we need the medium range structural order parameter captured through coarse-grained SOPs. This helps reconcile previous inconsistencies in static lengthscale growth observed in KALJ systems and supports the idea that dynamic slowdown is accompanied by the growth of structural order, albeit not purely local in nature.

In recent work, Tong and Tanaka have made an important and insightful observation regarding the common structural packing in systems with and without medium range crystalline order \cite{TongTanaka2018,tanakamrco_kawasaki_JPCM_2010}. It is well known that monodisperse or weakly polydisperse systems in two dimensions exhibit hexatic ordering of particles, which is directly related to the corresponding crystal structure. However, many supercooled liquid systems do not show any such ordering. Tong and Tanaka demonstrated that even in these systems, there exists a form of local ordering corresponding to the most sterically favored structure. Although this local order is not connected to any crystalline symmetry, it can be captured by their newly proposed order parameters, $\theta$ in two dimensions and $\Omega$ in three dimensions. This finding suggests that, for both systems with and without MRCO, the underlying tendency of a supercooled liquid is to achieve the most efficient local packing. In our study, the structural order parameter essentially characterizes the cage formed by the first-neighbor shell. This description naturally encodes information about local packing and is therefore similar in spirit to the order parameters introduced by Tong and Tanaka \cite{TongTanaka2018,TongTanaka2020}. Interestingly, and consistent with our present observations, it has been shown earlier that coarse-graining of $\theta$ and $\Omega$ successfully captures the growth of a static lengthscale even in MRCO-free systems, both in two and three dimensions \cite{TongTanaka2018,TongTanaka2020}.

Finally, we investigate whether the observed structural correlations are linked to underlying crystalline order. Previous analyses involving some of us showed that in the 80:20 mixture, the A particles exhibit a tendency to form fcc structures, while no evidence of CsCl-like ordering was observed~\cite{Ujjwal_jcp_2016}. This indicates that B particles do not participate in the crystallisation process in this composition. In contrast, the 60:40 mixture displays a tendency to form a CsCl structure, which involves both A and B particles~\cite{Ujjwal_jcp_2016}. Although in our simulations we do not see a crystallisation of the 60:40 mixture, we know that it is very close to the composition which crystallises,  i.e., the 58:42 mixture, as shown in Appendix V, crystallises primarily to a CsCl structure. Interestingly, neither the 80:20 nor the 60:40 composition exhibits growing locally favoured structures that correspond to the underlying crystalline order. Nevertheless, we find that the static lengthscale grows predominantly for the particle species associated with the dominant crystallisation tendency, namely, the $A$ particles in the 80:20 mixture, and both $A$ and $B$ particles in the 60:40 mixture. This suggests a potential link between the growth of structural correlations and latent crystallisation pathways. However, further investigations are needed to establish a definitive connection.

Overall, our study establishes a coherent framework that connects structural and dynamical perspectives in glass formation. It highlights the importance of choosing the correct spatial scale when analysing structural indicators relevant to the dynamical quantity. Future work could explore whether alternative structural indicators that have captured static lengthscale growth \cite{TongTanaka2019,Tanaka2025} in other systems can reveal similar signatures in the KALJ systems. Also, it will be important to test if other structural order parameters like the machine learning softness \cite{liu_nature,Cubuk2017} show growth in correlation length when coarse-grained.\\

\textbf{Appendix I: Temperature dependence of the distribution of the local SOP, $\beta \phi$}\\

\begin{figure}
    \centering
    \includegraphics[width=0.45\textwidth]{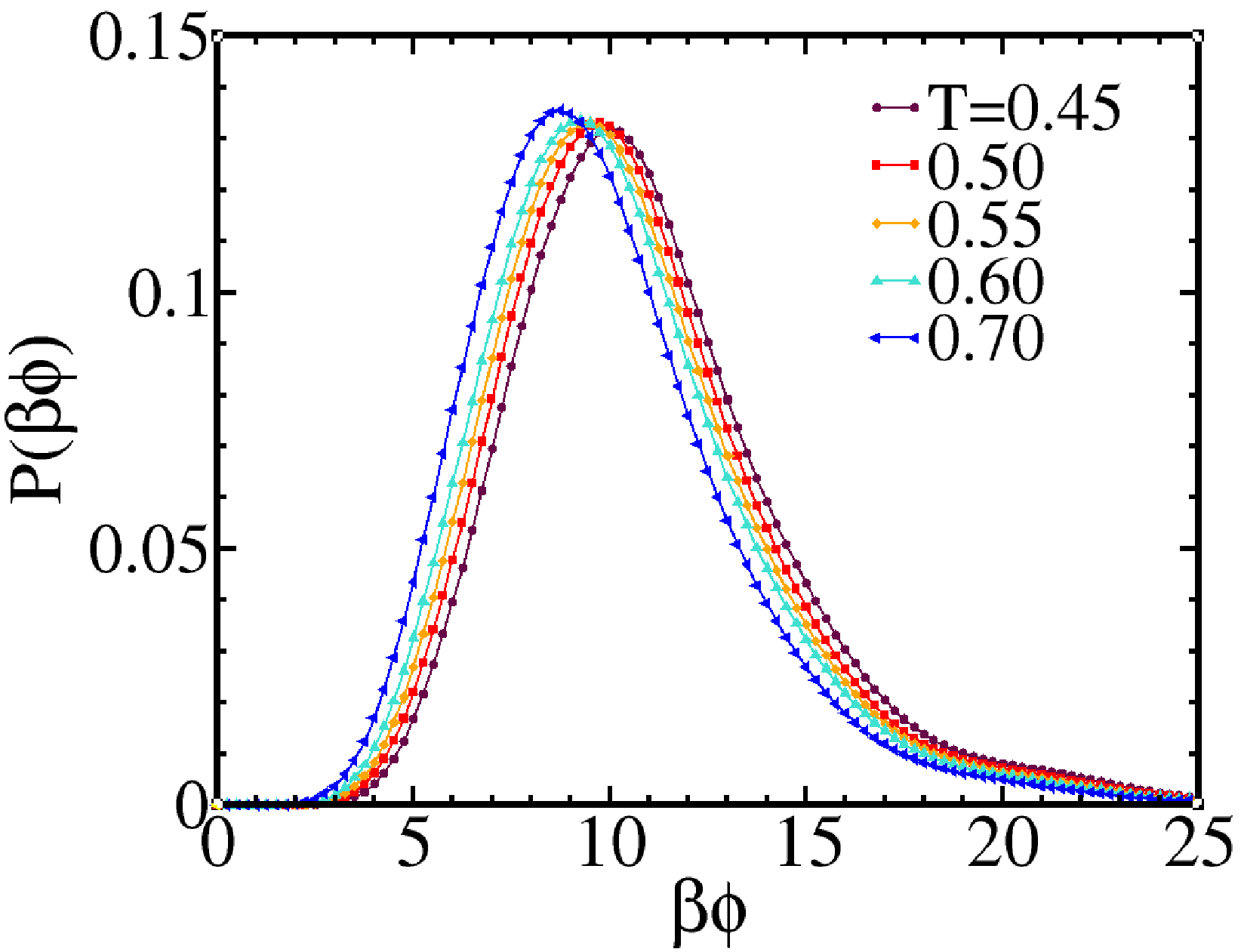}
        \caption{The temperature dependence of the distribution of the bare SOP, $\beta \phi$, for the 80:20 composition.}

    \label{fig:sopdistribution}
\end{figure}

The depth of the mean field caging potential, for a particle, \(\beta\phi\), is derived under the mean field approximation, which assumes a frozen background by neglecting the dynamics of the surrounding particles. We can compute it for every particle by using the information of the local rdf and the distribution \(P(\beta\phi)\) directly reflects spatial heterogeneity. With decreasing temperature, \(P(\beta\phi)\) shifts to larger values as shown in Fig. \ref{fig:sopdistribution}, indicating more particles in deeper cages. Thus \(\beta\phi\) captures the structural heterogeneity and its temperature evolution.\\

\textbf{Appendix II: Reliability of O.Z. Fitting in $S_4(q, t)$ }\\

\begin{figure}
    \centering
    \includegraphics[width=0.45\textwidth]{Fig11.eps}
        \caption{Scaling plot of $S_4(q, \tau_{\alpha})/S_4(q{=}0, \tau_{\alpha})$ versus $q\xi_4$. The dotted line represents the Ornstein-Zernike function used for fitting.}

    \label{fig:s4qt_reliable}
\end{figure}

In Fig.~\ref{fig:s4qt_reliable}, we show a scaling plot of $S_4(q, t)$ together with the Ornstein-Zernike (OZ) function (dotted line). The data exhibit good overlap with the OZ function at all temperatures, demonstrating that the fitting is reliable~\cite{Szamel_s4qt,Tah_dis_dis}.\\

\textbf{Appendix III: Dynamics in 80:20 and 60:40 composition in KALJ systems}\\

\begin{figure}
    \centering
    \includegraphics[width=0.45\textwidth]{Fig12.eps}
    \caption{The $\alpha$-relaxation time obtained from the overlap function (Eq.\ref{overlap}), $\tau_{\alpha}$, for two systems: 80:20 composition at $\rho=1.2$ and 60:40 composition at $\rho=1.41$ as a function of temperature. The solid lines are VFT fits, $\tau_{\alpha}=\tau_{o}*exp(\frac{T_{VFT}}{K_{VFT}(T-T_{VFT})})$ where $\tau_{o}$ is the prefactor, $K_{VFT}$ is the fragility which quantifies how fast the dynamics diverges, and  $T_{VFT}$ is the temperature where the dynamics is expected to diverge. For 80:20 mixture $T_{VFT}=0.319$ and for 60:40 mixture $T_{VFT}=0.312$}
    \label{fig:vft_fitting}
\end{figure}

In our study, we examine two KALJ systems at different compositions: the well-known 80:20 composition and a 60:40 composition whose density is tuned to $\rho=1.41$ so that both systems exhibited quantitatively similar relaxation dynamics in the same temperature range as shown in Fig.\ref{fig:vft_fitting}. Although the local structural motifs differ due to compositional variation, the overall relaxation behaviour is comparable in both systems. \\

\textbf{Appendix IV: Lengthscale studies in 80:20 composition}\\

\begin{figure}[H]
    \centering
    \includegraphics[width=0.4\textwidth]{Fig13.eps}
    \caption{
       In 80:20 mixture for A particles, the plot of the structural correlation length, $\xi_{\mathrm{CG}}$, versus the optimal coarse-graining lengthscale, $L_{\max}$, across all studied temperatures. The two measures exhibit a strong positive, near-linear relationship, indicating that  $\xi_{\mathrm{CG}}$ grows in tandem with the  $L_{\max}$.}
    \label{fig:xi_vs_Lmax}
\end{figure}

In Fig.\ref{fig:xi_vs_Lmax} we plot the structural correlation length, $\xi_{CG}$ against $L_{max}$, which shows that the former grows in tandem with the latter.

\begin{figure}
\centering
\begin{subfigure}{0.23\textwidth}
\caption{}
\includegraphics[width=0.75\linewidth]{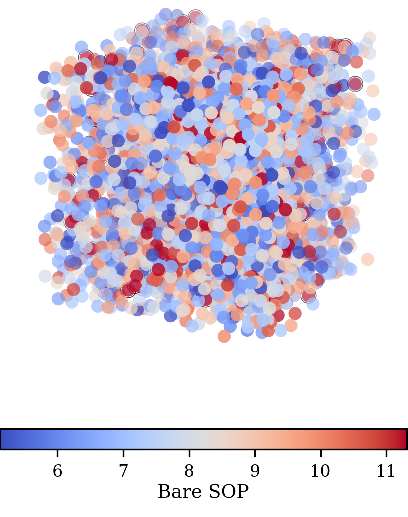}
\end{subfigure}
\begin{subfigure}{0.23\textwidth}
\caption{}
\vspace{0.5cm}
\includegraphics[width=1.1\linewidth]{Fig15b.eps}
\end{subfigure}
\begin{subfigure}{0.23\textwidth}
\caption{}
\includegraphics[width=0.75\linewidth]{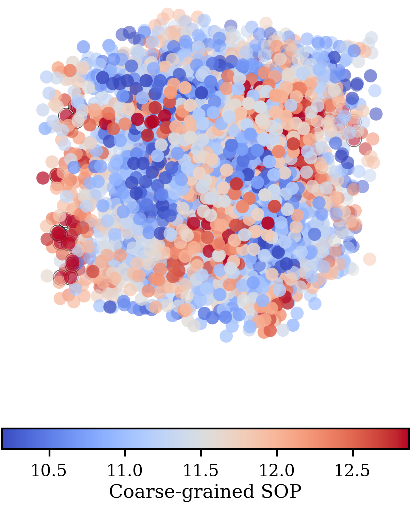}
\end{subfigure}
\begin{subfigure}{0.23\textwidth}
\caption{}
\vspace{0.5cm}
\includegraphics[width=1.1\linewidth]{Fig15d.eps}
\end{subfigure}
\begin{subfigure}{0.23\textwidth}
\caption{}
\includegraphics[width=0.75\linewidth]{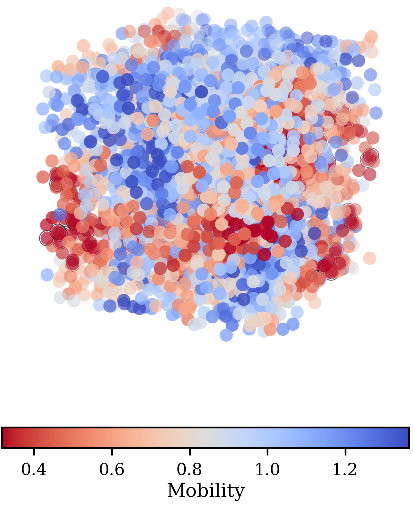}
\end{subfigure}
\begin{subfigure}{0.23\textwidth}
\caption{}
\vspace{0.5cm}
\includegraphics[width=1.1\linewidth]{Fig15f.eps}
\end{subfigure}
\caption{Snapshots of structural and dynamical properties and their corresponding excess correlations for B-type particles for 80:20 KALJ system at T=0.45. \textbf{(a)} Snapshot of the bare SOP, $\beta\phi$, appears spatially noisy, lacking visible large scale structure.  \textbf{(b)}The excess correlation of the bare SOP, $\Gamma_{\text{bare}}(r)$, (Eq.\ref{eq-gamma_static_bare}) confirms the absence of growth in bare static lengthscale.\textbf{(c)} Snapshot of the coarse-grained SOP, $\overline{\beta \phi}$, with $L_{max}$, shows emergence of some small patches of high and low SOP values. However these patches are much smaller than that observed for the A-type of particles (see Fig.\ref{fig:8020-correlation}(b)) \textbf{(d)}The excess correlation of the coarse-grained SOP, $\Gamma_{\text{CG}}(r)$, (Eq.\ref{eq-gamma_static_cg}) shows an increasing correlation length with decreasing temperature. Consistent with the size of the patches the static lengthscales obtained from the excess correlation are smaller than that of the A-type of particles (see Fig.\ref{fig:8020_compa_lengthscale})  \textbf{(e)}Snapshot of the particle mobility field, $\mu (\tau_{\alpha})$, calculated using Eq.\ref{eq:mobility} over the $\alpha$-relaxation time shows low and high mobility domains. \textbf{(f)} The excess displacement–displacement correlation function, $\Gamma_{uu}(r,t_\alpha)$, (Eq.\ref{eq_gamma}) exhibits growth with decreasing temperature.The colour coding in (d) and (f) matches that of (b).}
\label{fig:sop_Btype_8020}
\end{figure}

In Fig.\ref{fig:Cr-vs-time-B} we plot the Spearman rank correlation \( C_R(\overline{\beta\phi}, \mu) \) against the time scale by $\tau_{\alpha}$, which quantifies the relationship between the coarse-grained SOP and particle mobility for \( B \) particles. As shown in Fig.\ref{fig:Cr-vs-time-B}(a), for \( L = 0 \) (i.e., the bare SOP), the correlation is initially strong at short times but decays rapidly, indicating that the bare SOP captures structural features relevant only to short time dynamics. This decay becomes more pronounced at lower temperatures, as illustrated in Fig.\ref{fig:Cr-vs-time-B}(b). To predict dynamics over longer timescales, it becomes necessary to incorporate medium range structural information. This is evident from the increased structure–dynamics correlation when the SOP is coarse-grained over a spatial length \( L \); up to a certain optimum value, $L_{max}$, the coarse-grained SOP retains a stronger correlation with mobility at longer times.
Here, in Fig.\ref{fig:sop_Btype_8020}, we show the snapshots of the structural and dynamical parameters and their respective excess correlations for the B particles. We find that even for the coarse-grained SOP, the size of the patches representing high and low SOP values is smaller than that of the A particles (see Fig.\ref{fig:8020-correlation}(b)). This leads to the smaller values of $L_{max}$ and $\xi_{CG}$ for the B particles compared to the A particles.

\begin{figure}
  \centering
  \includegraphics[width=.4\textwidth]{Fig14a.eps}\par\vspace{6pt}
\vskip\baselineskip
  \includegraphics[width=.4\textwidth]{Fig14b.eps}
  \caption{Spearman rank correlation, $(C_{R}(\overline{\beta\Phi}, \mu))$, between the coarse-grained SOP, $\overline{\beta\phi}$, and mobility, \(\mu\), calculated at times scaled by the \(\alpha\)-relaxation time, \(t/\tau_{\alpha}\), for B-type particles at different coarse-graining lengths, \(L\). \textbf{(a)} Result at high temperature, \(T = 0.70\) and \textbf{(b)} at low temperature, \(T = 0.47\). The colour coding in (b) matches that of (a).}
  \label{fig:Cr-vs-time-B}
\end{figure}

\vspace{1cm}
\textbf{Appendix V: Crystallisation in KALJ 58:42  composition}\\

\begin{figure}
  \centering
  \includegraphics[width=.4\textwidth]{Fig16a.eps}\par\vspace{6pt}
\vskip\baselineskip
  \includegraphics[width=.4\textwidth]{Fig16b.eps}
\caption{\textbf{(a)} Potential energy, P.E., as a function of time, t,  at \( T = 0.66 \), showing a sudden drop indicative of a crystallisation event. We mark two regions before and after the drop as pre-crystalisation and post-crystalisation. \textbf{(b)} Probability distribution of the averaged bond orientational order parameter~\cite{Steinhardt1983,Mickel2013_boo} \(\overline{q}_6\)(Eq.\ref{q6}) for B-type particles with respect to A-type neighbours in the two regions {\it i.e} pre-crystalisation and post-crystalisation. A clear shift is observed from the pre-crystallisation state (black, circles) to the post-crystallisation state (red, squares), with the distribution moving to a \(\bar{q}_6\) value associated with bcc like local order. For comparison, the \(\overline{q}_6\) distribution for the 50:50 KALJ composition (green, triangle) which is known to crystallize into a CsCl structure~\cite{Atreyee_crystal_jcp} is also plotted.}
  \label{fig:58:42_crystal}
\end{figure}

In the 58:42 composition of the KALJ system, we observe clear signatures of crystallisation at a temperature \( T = 0.66 \). As shown in Fig.~\ref{fig:58:42_crystal}(a), the time evolution of the system’s potential energy exhibits a sudden drop, consistent with a crystallisation event. To further characterise this transition, we analysed the local bond orientational order using the averaged \(\overline{q}_l\) parameter for \( l = 6 \) given by~\cite{Steinhardt1983,Mickel2013_boo,Ujjwal_jcp_2016}:\\

\begin{equation}
\bar{q}_l = \sqrt{ \frac{4\pi}{2l + 1} \sum_{m=-l}^{l} \left| \bar{q}_{lm} \right|^2 }
\label{q6}
\end{equation}

\noindent
where, $\bar{q}_{lm}(i)=\frac{1}{{N}_i} \sum_{k=0}^{{N}_i} q_{lm}(k)$ and $q_{lm}(i)=\frac{1}{N_i} \sum_{j=1}^{N_i} Y_{lm} \left( \theta(r_{ij}), \phi(r_{ij}) \right)$. Here, \( Y_{lm} \) are spherical harmonics, and \( N_i \) is the number of neighbors of particle \( i \) within a cutoff radius \( r_{\text{cut}} = 1.2 \). The quantity \( \bar{q}_6 \) characterises the degree of six-fold local orientational symmetry, averaged over the neighbourhood of particle \( i \). In Fig.~\ref{fig:58:42_crystal}(b), we present the distribution of \( \overline{q}_6 \) values computed for B-type particles with respect to their neighbouring A-type particles. Notably, after the energy drop in Fig.\ref{fig:58:42_crystal}(b), the distribution of BOO shifts to the range which is characteristic of bcc symmetry ($\bar{q_6} = 0.6 $). 
This behaviour closely correlates with the local structural environment found in the well-established CsCl-type crystalline phase, typically observed in the equimolar (50:50) KALJ composition at low temperatures~\cite{Atreyee_crystal_jcp}. The similarity in the \( \overline{q}_6 \) distributions between the 58:42 and 50:50 compositions further supports the presence of CsCl ordering in the 58:42 KALJ system.

\vspace{1cm}

\begin{acknowledgments}
S.~M.~B. thanks, Science and Engineering Research Board (SERB, Grant No. SPF/2021/000112 ) and CSIR-National Chemical Laboratory (NCL, Grant No. MLP046726) for funding. S.~K. thanks DST INSPIRE for Fellowship. M.~S. thanks CSIR for the fellowship. S.~M.~B. would like to thank Smarajit Karmakar, Chandan Dasgupta and Patrick Royall for discussions. S.~K. would like to thank Palak Patel and Anoop Mutneja for discussions.\\   
\end{acknowledgments}

{\bf AVAILABILITY OF DATA}\\
The data that support the findings of this study are available from the corresponding author upon reasonable request.\\[3mm]


\vspace{1cm}
\section{REFERENCES}
\begin{thebibliography}{10}

\bibitem{Debenedetti2001}
P.~G. Debenedetti and F.~H. Stillinger,
\newblock Nature {\bf 410}, 259 (2001).

\bibitem{Cavagna2009}
A.~Cavagna,
\newblock Physics Reports {\bf 476}, 51 (2009).

\bibitem{Berthier2011}
L.~Berthier and G.~Biroli,
\newblock Reviews of Modern Physics {\bf 83}, 587 (2011).

\bibitem{Ediger2000}
M.~D. Ediger,
\newblock Annual Review of Physical Chemistry {\bf 51}, 99 (2000).

\bibitem{tanakamrco_kawasaki_prl_2007}
T.~Kawasaki, T.~Araki, and H.~Tanaka,
\newblock Physical Review Letters {\bf 99}, 215701 (2007).

\bibitem{Coslovich2007}
D.~Coslovich and G.~Pastore,
\newblock The Journal of Chemical Physics {\bf 127}, 124504 (2007).

\bibitem{royall2008}
C.~P. Royall, S.~R. Williams, T.~Ohtsuka, and H.~Tanaka,
\newblock Nature Materials {\bf 7}, 556 (2008).

\bibitem{tanakamrco_shintani_natphys_2008}
H.~Shintani and H.~Tanaka,
\newblock Nature Materials {\bf 7}, 870 (2008).

\bibitem{Karmakar2009}
S.~Karmakar, C.~Dasgupta, and S.~Sastry,
\newblock Proceedings of the National Academy of Sciences {\bf 106}, 3675 (2009).

\bibitem{tanakamrco_critical_natmat_2010}
H.~Tanaka, T.~Kawasaki, H.~Shintani, and K.~Watanabe,
\newblock Nature Materials {\bf 9}, 324 (2010).

\bibitem{keys2011}
A.~S. Keys, L.~O. Hedges, J.~P. Garrahan, S.~C. Glotzer, and D.~Chandler,
\newblock Physical Review X {\bf 1}, 021013 (2011).

\bibitem{Coslovich2011}
D.~Coslovich,
\newblock The Journal of Chemical Physics {\bf 134}, 034504 (2011).

\bibitem{Sastry2014}
S.~Sastry,
\newblock Annual Review of Condensed Matter Physics {\bf 5}, 129 (2014).

\bibitem{bapst2020}
V.~Bapst {\em et~al.},
\newblock Proceedings of the National Academy of Sciences {\bf 117}, 13201 (2020).

\bibitem{Manoj_prl_2021}
M.~K. Nandi and S.~M. Bhattacharyya,
\newblock Physical Review Letters {\bf 126}, 208001 (2021).

\bibitem{Sharma2022}
M.~Sharma, M.~K. Nandi, and S.~M. Bhattacharyya,
\newblock Physical Review E {\bf 105}, 044604 (2022).

\bibitem{shintani2006}
H.~Shintani and H.~Tanaka,
\newblock Nature Physics {\bf 2}, 200 (2006).

\bibitem{Dunleavy2012}
A.~J. Dunleavy, K.~Wiesner, and C.~P. Royall,
\newblock Physical Review E {\bf 86}, 041505 (2012).

\bibitem{Tarjus_simpleglass_2012}
B.~Charbonneau, P.~Charbonneau, and G.~Tarjus,
\newblock Physical Review Letters {\bf 108}, 035701 (2012).

\bibitem{MalinsLifetimes2013}
A.~Malins, J.~Eggers, H.~Tanaka, and C.~P. Royall,
\newblock Faraday Discussions {\bf 167}, 405 (2013).

\bibitem{russo_tanaka}
J.~Russo and H.~Tanaka,
\newblock Proceedings of the National Academy of Sciences {\bf 112}, 6920 (2015).

\bibitem{Tah_PRL_2018}
I.~Tah, S.~Sengupta, S.~Sastry, C.~Dasgupta, and S.~Karmakar,
\newblock Physical Review Letters {\bf 121}, 085703 (2018).

\bibitem{Karmakar2010}
S.~Karmakar, C.~Dasgupta, and S.~Sastry,
\newblock Physical Review Letters {\bf 105}, 015701 (2010).

\bibitem{Tah_dis_dis}
I.~Tah and S.~Karmakar,
\newblock Physical Review Research {\bf 2}, 022067 (2020).

\bibitem{Paula2023}
K.~Paula, A.~Mutneja, S.~K. Nandi, and S.~Karmakar,
\newblock Proceedings of the National Academy of Sciences {\bf 120}, e2217073120 (2023).

\bibitem{Donati1999PRL}
C.~Donati, S.~C. Glotzer, and P.~H. Poole,
\newblock Physical Review Letters {\bf 82}, 5064 (1999).

\bibitem{Flenner2013}
E.~Flenner and G.~Szamel,
\newblock The Journal of Chemical Physics {\bf 138}, 12A523 (2013).

\bibitem{Hallett2018}
J.~E. Hallett, F.~Turci, and C.~P. Royall,
\newblock Nature Communications {\bf 9}, 3272 (2018).

\bibitem{berthier_PRX_2022}
C.~Scalliet, B.~Guiselin, and L.~Berthier,
\newblock Physical Review X {\bf 12}, 041028 (2022).

\bibitem{Steinhardt1983}
P.~J. Steinhardt, D.~R. Nelson, and M.~Ronchetti,
\newblock Physical Review B {\bf 28}, 784 (1983).

\bibitem{tanakamrco_kawasaki_JPCM_2010}
T.~Kawasaki and H.~Tanaka,
\newblock Journal of Physics: Condensed Matter {\bf 22}, 232102 (2010).

\bibitem{tanakamrco_leocmach_natcomm_2012}
M.~Leocmach and H.~Tanaka,
\newblock Nature Communications {\bf 3}, 974 (2012).

\bibitem{tanakamrco_boo_jncs_2012}
H.~Tanaka,
\newblock Journal of Non-Crystalline Solids {\bf 351}, 3371 (2012).

\bibitem{tanakamrco_hu_pree_2016}
Y.-C. Hu, H.~Tanaka, and W.-H. Wang,
\newblock Physical Review E {\bf 94}, 042602 (2016).

\bibitem{TongTanaka2018}
H.~Tong and H.~Tanaka,
\newblock Physical Review X {\bf 8}, 011041 (2018).

\bibitem{TarjusPRL2010}
F.~Sausset and G.~Tarjus,
\newblock Physical Review Letters {\bf 104}, 065701 (2010).

\bibitem{biroli_thermodynamic_2008}
G.~Biroli, J.-P. Bouchaud, A.~Cavagna, T.~S. Grigera, and P.~Verrocchio,
\newblock Nature Physics {\bf 4}, 771 (2008).

\bibitem{Hocky2012}
G.~M. Hocky, T.~E. Markland, and D.~R. Reichman,
\newblock Physical Review Letters {\bf 108}, 225506 (2012).

\bibitem{Biroli2013}
G.~Biroli, S.~Karmakar, and I.~Procaccia,
\newblock Physical Review Letters {\bf 111}, 165701 (2013).

\bibitem{Berthier2016}
L.~Berthier, P.~Charbonneau, and S.~Yaida,
\newblock The Journal of Chemical Physics {\bf 144}, 024501 (2016).

\bibitem{Marinari2005}
E.~Marinari and E.~Pitard,
\newblock Europhysics Letters {\bf 69}, 235 (2005).

\bibitem{LaNave2006}
E.~La~Nave, S.~Sastry, and F.~Sciortino,
\newblock Physical Review E {\bf 74}, 050501(R) (2006).

\bibitem{Malins2013JCP}
A.~Malins, S.~R. Williams, J.~Eggers, and C.~P. Royall,
\newblock The Journal of Chemical Physics {\bf 139}, 234506 (2013).

\bibitem{Royall2014}
C.~P. Royall, A.~Malins, A.~J. Dunleavy, and R.~Pinney,
\newblock Journal of Non-Crystalline Solids {\bf 407}, 34 (2014).

\bibitem{Crowther2015}
P.~Crowther, F.~Turci, and C.~P. Royall,
\newblock The Journal of Chemical Physics {\bf 143}, 044503 (2015).

\bibitem{KARMAKAR20121001}
S.~Karmakar, E.~Lerner, and I.~Procaccia,
\newblock Physica A: Statistical Mechanics and its Applications {\bf 391}, 1001 (2012).

\bibitem{Chakrabarty2017}
S.~Chakrabarty, I.~Tah, S.~Karmakar, and C.~Dasgupta,
\newblock Physical Review Letters {\bf 119}, 205502 (2017).

\bibitem{FernandezHarrowell_PRE_2003}
J.~R. Fernández and P.~Harrowell,
\newblock Physical Review E {\bf 67}, 011403 (2003).

\bibitem{Ujjwal_jcp_2016}
U.~K. Nandi, A.~Banerjee, S.~Chakrabarty, and S.~M. Bhattacharyya,
\newblock The Journal of Chemical Physics {\bf 145}, 034502 (2016).

\bibitem{liu_nature}
S.~S. Schoenholz, E.~D. Cubuk, D.~M. Sussman, E.~Kaxiras, and A.~J. Liu,
\newblock Nature Physics {\bf 12}, 469 (2016).

\bibitem{Zhang2020}
Z.~Zhang and W.~Kob,
\newblock Proceedings of the National Academy of Sciences {\bf 117}, 14032 (2020).

\bibitem{TongTanaka2019}
H.~Tong and H.~Tanaka,
\newblock Nature Communications {\bf 10}, 5596 (2019).

\bibitem{TongTanaka2020}
H.~Tong and H.~Tanaka,
\newblock Physical Review Letters {\bf 124}, 225501 (2020).

\bibitem{Tanaka2025}
H.~Tanaka,
\newblock The Journal of Physical Chemistry B {\bf 129}, 789 (2025).

\bibitem{ken_sch}
B.~Mei and K.~S. Schweizer,
\newblock The Journal of Physical Chemistry B {\bf 128}, 11293 (2024).

\bibitem{MeiWang2022}
B.~Mei, B.~Zhuang, Y.~Lu, L.~An, and Z.-G. Wang,
\newblock The Journal of Physical Chemistry Letters {\bf 13}, 3957 (2022).

\bibitem{mohit_wca}
M.~Sharma, M.~K. Nandi, and S.~M. Bhattacharyya,
\newblock The Journal of Chemical Physics {\bf 159}, 104502 (2023).

\bibitem{QiuArun2025}
Y.~Qiu, I.~Jang, X.~Huang, and A.~Yethiraj,
\newblock Proceedings of the National Academy of Sciences {\bf 122}, e2427246122 (2025).

\bibitem{Kob1994}
W.~Kob and H.~C. Andersen,
\newblock Physical Review Letters {\bf 73}, 1376 (1994).

\bibitem{Stoddard1973}
S.~D. Stoddard and J.~Ford,
\newblock Physical Review A {\bf 8}, 1504 (1973).

\bibitem{Poole1998PhysicaA}
P.~H. Poole, C.~Donati, and S.~C. Glotzer,
\newblock Physica A {\bf 261}, 51 (1998).

\bibitem{LandauSchuttler1999}
D.~P. Landau and H.-B. Sch\"{u}ttler, editors,
\newblock {\em Computer Simulation Studies in Condensed-Matter Physics XI}, Springer Proceedings in Physics Vol.~84 (Springer-Verlag, Berlin Heidelberg, 1999).

\bibitem{RamakrishnanYussouff1979}
T.~V. Ramakrishnan and M.~Yussouff,
\newblock Physical Review B {\bf 19}, 2775 (1979).

\bibitem{HansenMcDonald2013}
J.-P. Hansen and I.~R. McDonald,
\newblock {\em Theory of Simple Liquids}, 4 ed. (Academic Press, Oxford, 2013).

\bibitem{PiaggiValssonParrinello2017}
P.~M. Piaggi, O.~Valsson, and M.~Parrinello,
\newblock Physical Review Letters {\bf 119}, 015701 (2017).

\bibitem{PatelSharmaBhattacharyya2023}
P.~Patel, M.~Sharma, and S.~M. Bhattacharyya,
\newblock The Journal of Chemical Physics {\bf 159}, 044501 (2023).

\bibitem{Cubuk2017}
E.~D. Cubuk {\em et~al.},
\newblock Science {\bf 358}, 1033 (2017).

\bibitem{harrowell}
A.~Widmer-Cooper and P.~Harrowell,
\newblock Journal of Physics: Condensed Matter {\bf 17}, S4025 (2005).

\bibitem{Szamel_s4qt}
E.~Flenner and G.~Szamel,
\newblock Physical Review Letters {\bf 105}, 217801 (2010).

\bibitem{Berthier_Biroli_PRE_2005}
C.~Toninelli, M.~Wyart, L.~Berthier, G.~Biroli, and J.-P. Bouchaud,
\newblock Physical Review E {\bf 71}, 041505 (2005).

\bibitem{das_possible_2018}
R.~Das, I.~Tah, and S.~Karmakar,
\newblock The Journal of Chemical Physics {\bf 149}, 024501 (2018).

\bibitem{Tanaka2012bondliquid}
H.~Tanaka,
\newblock European Physical Journal E {\bf 35} (2012).

\bibitem{Langer2013}
J.~S. Langer,
\newblock Physical Review E {\bf 88}, 012122 (2013).

\bibitem{Lubchenko2007}
V.~Lubchenko and P.~G. Wolynes,
\newblock Annual Review of Physical Chemistry {\bf 58}, 235 (2007).

\bibitem{Kirkpatrick1989}
T.~R. Kirkpatrick, D.~Thirumalai, and P.~G. Wolynes,
\newblock Physical Review A {\bf 40}, 1045 (1989).

\bibitem{Royall_excitations_2025}
D.~Lang, C.~Scalliet, and C.~P. Royall,
\newblock Physical Review E {\bf 111}, 055415 (2025).

\bibitem{Yaida2016}
P.~Charbonneau, E.~Dyer, J.~Lee, and S.~Yaida,
\newblock Journal of Statistical Mechanics: Theory and Experiment {\bf 2016}, 074004 (2016).

\bibitem{Mickel2013_boo}
W.~Mickel, S.~C. Kapfer, G.~E. Schröder-Turk, and K.~Mecke,
\newblock The Journal of Chemical Physics {\bf 138}, 044501 (2013).

\bibitem{Atreyee_crystal_jcp}
A.~Banerjee, S.~Chakrabarty, and S.~M. Bhattacharyya,
\newblock The Journal of Chemical Physics {\bf 139}, 104501 (2013).

\end{thebibliography}
\end{document}